\documentclass[twocolumn,showpacs,floatfix,superscriptaddress,amsmath,amssymb,prl]{revtex4}
\usepackage{mathrsfs}
\usepackage{txfonts}
\usepackage{amssymb}
\usepackage{graphicx}
\usepackage{hyperref}
\usepackage{ulem}
\usepackage{overpic}

\usepackage{color}
 \usepackage{psfrag}
\usepackage{tabularx}
\usepackage{array}
\usepackage{placeins}
\makeatletter

\begin{document}

\title{Absence of a critical nematic phase in the vicinity of the $\rm {SU}(3)$ ferromagnetic point for the one-dimensional spin-1 bilinear-biquadratic model}

\author{Yan-Wei Dai}
\affiliation{Centre for Modern Physics,
Chongqing University, Chongqing 400044, The People's Republic of
China}

\author{Qian-Qian Shi}
\affiliation{Centre for Modern Physics,
Chongqing University, Chongqing 400044, The People's Republic of
China}

\author{Huan-Qiang Zhou}
\affiliation{Centre for Modern Physics,
Chongqing University, Chongqing 400044, The People's Republic of
China}

\author{Ian P. McCulloch}
\affiliation{ Department of Physics, National Tsing Hua University, Hsinchu 30013, Taiwan}
\affiliation{School of Mathematics and Physics, The University of Queensland, St. Lucia, QLD 4072, Australia}
\affiliation{Centre for Modern Physics, Chongqing University, Chongqing 400044, The People's Republic of China}

\begin{abstract}
The absence of a critical nematic phase in the vicinity of the $\rm {SU}(3)$ ferromagnetic point for the one-dimensional spin-1 bilinear-biquadratic model is demonstrated by means of the tensor network algorithms.  As it turns out, the phase transition from the ferromagnetic phase to the dimerized phase at the $\rm {SU}(3)$ ferromagnetic point is direct, but not of the first-order.  The transition point features highly degenerate ground states, which are scale but not conformally invariant, with the fractal dimension being equal to 2.  The conceptual developments in effective field theories - the fractal dimension and the counting rule of the Goldstone modes - play a pivotal role in clarifying the numerical artifacts arising from the finiteness of the bond dimension in the tensor network simulations, which are attributed to a proximity effect to a highly entangled scale or conformally invariant ground state.

\end{abstract}
\maketitle
\begin{figure}
\includegraphics[width=0.25\textwidth]{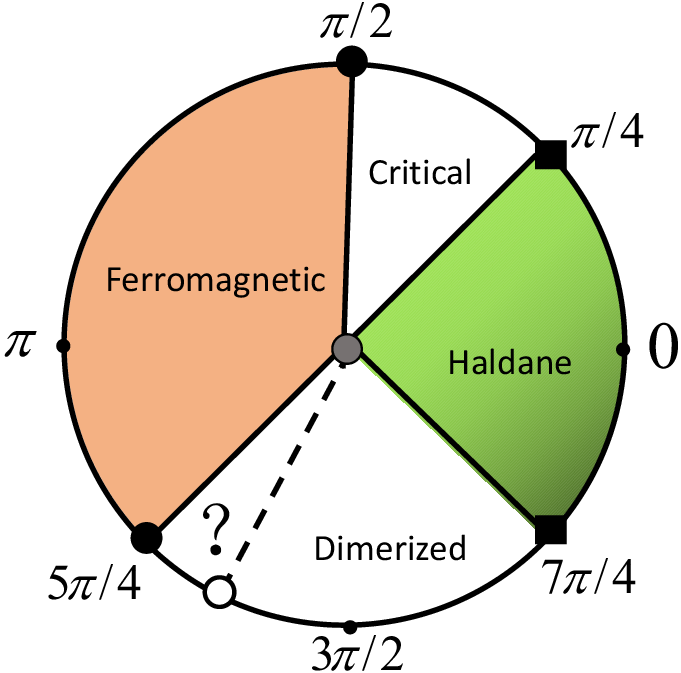}
\caption{(color online) A sketch of the ground state phase diagram for the one-dimensional spin-1 bilinear-biquadratic model. Here, $\phi$
 ranges from 0 to $2\pi$, but we focus on a region in the vicinity of the $\rm{SU}(3)$ ferromagnetic point at $\phi = 5/4 \pi$.
The question mark indicates a possible critical nematic phase, which turns out to be absent.
 } \label{iFig1}
\end{figure}

{\it Introduction.-}
The spin-1 bilinear-biquadratic model has been investigated extensively, either analytically or
numerically~\cite{Chubukov, Fath2, Kawashima, Ivanov,  Buchta, Rizzi, Lauchli, Porras, Romero, Rakov, Sierra, Fath,ronny}.
It is described
by the Hamiltonian with the nearest-neighbor interactions

\begin{equation}
 H = \sum_{j}
 \left[\cos\phi (\mathbf{S}_{j}\mathbf{S}_{j+1})
       +\sin\phi (\mathbf{S}_{j}\mathbf{S}_{j+1})^2
       \right],
 \label{ham1}
\end{equation}
where $\mathbf{S}_j=(S_{j}^{x},S_{j}^{y},S_{j}^{z})$ denote the spin-1 operators acting on the $j$-th site.
The model exhibits rich physics, with the ground-state phase diagram being shown in Fig.~\ref{iFig1}.
A remarkable example is the Haldane gap in
the spin-1 Heisenberg antiferromagnet~\cite{Haldane} at $\phi=0$, which is adiabatically connected to the
Affleck-Kennedy-Lieb-Tasaki point at  $\phi=\arctan~(1/3)$~\cite{AKLT}, with its ground state being the valence-bond-solid state.
This exactly solved point provides compelling evidence
for the existence of the gapped Haldane phase with hidden topological order~\cite{Haldane,Oshikawa} for
$-\pi/4<\phi<\pi/4$. The importance of the model (\ref{ham1}) is also embodied in the fact that it is exactly solvable by means of
the Bethe ansatz at a few selected points, including $\phi=\pi/4$~\cite{Sutherland}, $\phi=-\pi/2$~\cite{TB}, and $\phi=-\pi/4$~\cite{Barber}.
In addition, the model exhibits the $\rm {SU}(3)$ symmetry~\cite{Batista,Affleck,chenxh}
at $\phi=\pm\pi/2$, $\pi/4$, and $-3\pi/4$
(cf. also Sec.~SI in the Supplementary Material (SM)).
However, whether or not the phase transition from the ferromagnetic phase to the dimerized phase is direct
remains to be controversial up to present.

The controversy dates back to a pioneering work by Chubukov~\cite{Chubukov}, who suggested that a gapped nematic phase exists between the dimerized phase and the ferromagnetic phase. In contrast, F\'{a}th and S\'{o}lyom~\cite{Fath2} advocated that the phase transition from the ferromagnetic phase to the dimerized phase is direct and of the first-order, thus ruling out the existence of the gapped nematic phase.
They also argued that an exponentially decaying small gap opens once away from the $\rm{SU}(3)$ ferromagnetic point. However, no mechanism is offered to account for the opening of such an exponentially decaying small gap.
Subsequently, an extensive investigation has been done~\cite{Kawashima,Ivanov,Buchta,Rizzi,Lauchli,Porras,Romero,Rakov}, in attempt to settle this controversial issue.  Buchta \textit{et al}~\cite{Buchta} showed that the dimerized phase
prevails down to the $\rm{SU}(3)$ point, although a non-dimerized phase is possible in a
narrow range close to the $\rm{SU}(3)$ ferromagnetic point. In Refs.~\cite{Lauchli, Rakov}, a critical nematic phase is suggested as a
possible non-dimerized phase between the ferromagnetic phase and the dimerized phase. Meanwhile,
Rizzi \textit{et al}~\cite{Rizzi} concluded
that there is no intermediate nematic phase, but a tendency is enhanced towards the nematic order.
Instead, L\"{a}chli, Schmid, and
Trebst~\cite{Lauchli} suggested that
an unconventional crossover might lie at the heart of the complications concerning the presence or absence of a
critical nematic phase.
In a recent work~\cite{Rakov},
the $\rm{SU}(2)$ symmetry is implemented in a finite-size matrix product state (MPS) algorithm under the periodic boundary conditions, which is sufficient to confirm the absence of a critical nematic phase for
$\phi/\pi>-0.72$, but leaving it open the possibility for
the existence of a critical nematic phase in the vicinity of the $\rm{SU}(3)$ ferromagnetic point.

In this work, we adopt a completely different strategy to address this controversy. In our opinion, the current
state-of-the-art algorithms in the context of the tensor network representations are powerful enough to efficiently simulate the model (\ref{ham1}), including
the infinite time evolving block decimation (iTEBD)~\cite{Vidal}, the ${\rm U}(1)$ infinite density matrix renormalization group (iDMRG) algorithm, and the $\rm{SU}(2)$
iDMRG algorithm~\cite{mcculloch}. For our purpose, we focus on the region $5\pi/4<\phi<3\pi/2$.
As it turns out, different types of numerical artifacts arise from the finiteness of the bond
dimension in the tensor network simulations, if different symmetries are implemented.
The artifacts may be characterized in terms of the fractal dimension~\cite{Doyon} and the counting rule of the Goldstone modes (GMs)~\cite{Watanabe}
- recent conceptual developments in effective field theories. As a consequence,
there are two distinct types of the artifacts, which may be attributed to a proximity effect to two types of highly entangled ground states: one is conformally invariant and the other is scale invariant, thus leading to a pseudo critical regime and a pseudo fractal regime. Here, by ``pseudo" we mean an artifact arising from the finiteness of the bond dimension in the tensor network simulations.
It is the presence of the numerical artifacts that
makes it very hard to extract the underlying physics behind the model (\ref{ham1}).

\begin{figure}
\includegraphics[width=0.37\textwidth]{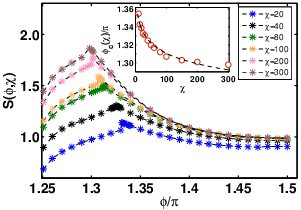}
\caption{(color online) Main: the entanglement entropy $S(\phi,\chi)$ as a function of $\phi$,
 with the bond dimension $\chi= 20,40,80,100,200$ and $300$, from the iTEBD simulation. One pseudo critical point $\phi_{c}(\chi)$ is detected as a peak for each value of the bond dimension $\chi$.
 Inset: an extrapolation of the pseudo critical
points $\phi_{c}(\chi)$ is performed, as the bond dimension $\chi$ tends to infinity. Here, the termination point is chosen to be
at the $\rm{SU}(3)$ ferromagnetic point. Thus, the data is fitted in the form:
$\phi_{c}(\chi)= 5/4 \pi+ p \chi ^q$, where  $p=-2.54$ and $q=-0.0064$.
 } \label{iFig2}
\end{figure}


{\it The $\rm{iTEBD}$ simulation.-} The iTEBD simulation is performed in the
region $(1.25\pi, 1.5\pi]$. In Fig.~\ref{iFig2}, the entanglement entropy $S(\phi,\chi)$, which quantifies the entanglement of a bipartite system~\cite{Bennett}, is plotted as a function
of $\phi$, with the bond dimension $\chi= 20,40,80,100,200$ and $300$.
Here, we note that
in the infinite MPS representation, the entanglement entropy $S(\phi,\chi)$ for the semi-infinite chain may be written as
$S(\phi,\chi)=-\sum_{\mu}\lambda_{\mu}^2(\phi,\chi)\ln\lambda_{\mu}^2(\phi,\chi)$, with
$\lambda_\mu^2(\phi,\chi)$ being the Schmidt decomposition coefficients.
The entanglement entropy $S(\phi,\chi)$ exhibits a singular peak, with each singular
point being indicative of a pseudo critical point $\phi_c(\chi)$.
Hence, there is a pseudo critical point  $\phi_c(\chi)$ for a given value of the bond dimension $\chi$,
separating the whole region into a pseudo critical regime and a dimerized regime.
As an illustrative example,
for the bond dimension $\chi=300$, the pseudo critical point $\phi_{c}(\chi)$ is located at $\phi_{c}(\chi)=1.298\pi$.

In order to characterize the pseudo critical regime, we need to extract the central charge $c$ from the finite-entanglement scaling~\cite{Tagliacozzo, Pollmann}:
 \begin{equation}
            S(\phi,\chi)= \frac{c}{6}\ln{\xi(\phi,\chi)}+a(\phi),
            \label{fes}
 \end{equation}
where $\xi(\phi,\chi)\propto\chi^{\kappa(\phi)}$, with $\kappa(\phi)$ being the finite entanglement scaling exponent, and $a(\phi)$ is an additive constant.
The best linear fit is performed for $\phi=1.26\pi,1.27\pi,1.28\pi$ and $1.29\pi$,
with the bond dimension ranging from $16$ to $300$, respectively.
The central charge $c$ is estimated to be $c=2$, with a relative error being less than $2\%$, in the entire
pseudo critical regime (for details, see Sec.~SII in SM).
Meanwhile, the central charge $c$ is also extracted from the pseudo critical points $\phi_{c}(\chi)$, yielding $c=1.86$,
with a relative error being less than $7\%$, compared to the expected exact value $c=2$ (cf. Sec.~SII in SM). We attribute a higher relative error in this case to the fact that the
accuracies of the pseudo critical points $\phi_c(\chi)$ are relatively lower, due to high computational costs.

Physically, the fact that the central charge $c=2$ in the pseudo critical regime may be explained in terms of the counting rule of the (pseudo) GMs adapted to numerical artifacts.  Actually, this arises from the pseudo spontaneous symmetry breaking (SSB)~\cite{Wang}
from $\rm{SU}(2)$  to ${\rm U}(1)$,
in which two generators are spontaneously broken, thus resulting in two pseudo GMs of type-A, with the central charge $c$ being the number of the pseudo GMs, given that the central charge $c$ measures the number of the gapless excitations~\cite{CFT}. Here, we note that the infinite MPS algorithm naturally leads to infinitely degenerate ground states in a pseudo critical regime, due to the finiteness of the
bond dimension. In fact, if the bond dimension $\chi$ tends to infinity, then a pseudo local
order parameter $\langle O_j \rangle$ arising from pseudo SSB tends to vanish, as required to keep consistency
with the Mermin-Wagner theorem~\cite{Mermin}.
In our case, the pseudo local order parameter $\langle O_j \rangle$ is defined as follows
 \begin{equation}
             {\langle O_j \rangle=\sqrt{\sum^8_{\alpha,\;\beta =1}g^{\alpha \beta}\langle K^j_{\alpha}\rangle \langle K^j_{\beta}\rangle}},
 \label{localorder}
 \end{equation}
where $g^{\alpha \beta}=g_{\alpha \beta}^{-1}$, $g_{\alpha \beta}$ is a metric tensor: $g_{\alpha\beta}=\sum_{\delta,\epsilon}\gamma_{\alpha\delta\epsilon}\gamma_{\beta\delta\epsilon}$ ($\alpha, \beta, \delta,\epsilon = 1, 2, \cdots, 8$),
with $\gamma_{\alpha\beta\delta}$ being the structural constants of the $\rm{SU}(3)$ group,
and $K^j_{\alpha}$ is the local components of the $\rm{SU}(3)$ generators  at the $j$-th site.
Since the iTEBD simulation is performed for a randomly chosen initial state, it yields a ground state wave
function in the infinite MPS representation, with
different expectation values of the generators $K^j_{\alpha}$. However, the pseudo local order operator $\langle O_j \rangle$ does
not depend on an initial state, within accuracies.  It is found that
the same set of the pseudo critical points $\phi_{c}(\chi)$ are detected in terms of the pseudo local order parameter
 $\langle O_j (\chi)\rangle$,
consistent with those from the entanglement entropy $S(\phi,\chi)$
and the pseudo local order parameter $\langle O_j(\chi) \rangle$ is scaled down to zero (for details, cf Sec.~SIV in SM).

In addition, the dimerized phase is characterized in terms of the local order parameter
$\langle D_{j,j+1} \rangle=\langle \mathbf{S}_{j}\mathbf{S}_{j+1}-\mathbf{S}_{j+1}\mathbf{S}_{j+2} \rangle$,
which tends to be saturated as the bond dimension $\chi$ increases.

A crucial question concerns whether or not the pseudo critical points $\phi_{c}(\chi)$ terminate at the $\rm{SU}(3)$ ferromagnetic point, if the bond dimension $\chi$ tends to infinity. As shown in Fig.~\ref{iFig2}, it is possible for
the pseudo critical
points $\phi_{c}(\chi)$  to terminate at the $\rm{SU}(3)$ ferromagnetic point. Here, the termination point is chosen to be
at the $\rm{SU}(3)$ ferromagnetic point. Thus, the fitting function takes the form:
$\phi_{c}(\chi)= 5/4 \pi+ p \chi ^q$, where  $p=-2.54$ and $q=-0.0064$.
However, such an extrapolation also works within the accuracies, if we  make any other choice of the termination point, as long as it is close enough to the $\rm{SU}(3)$ ferromagnetic point.
As a consequence, two possible scenarios arise:
(i) the pseudo critical points $\phi_c(\chi)$ terminate at the $\rm{SU}(3)$ ferromagnetic point, when the bond dimension $\chi$ tends to infinity;
(ii) the pseudo critical points $\phi_c(\chi)$ terminate at a point $\phi_c$ away from the $\rm{SU}(3)$ ferromagnetic point, when the bond
dimension $\chi$ tends to infinity.
In the first scenario, there is no critical nematic phase between the $\rm{SU}(3)$ ferromagnetic point and the dimerized phases.
In the second scenario, there is a critical nematic phase between the $\rm{SU}(3)$ ferromagnetic point and the dimerized phases,
with the central charge $c$ being $2$ (for details, cf Sec.~SVI in SM).

\begin{figure}
\includegraphics[width=0.37\textwidth]{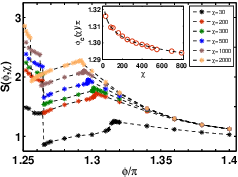}
\caption{(color online) Main: the entanglement entropy $S(\phi,\chi)$ as a function of $\phi$,
 with the bond dimension $\chi=30,200,300,500,1000$ and $2000$, from the ${\rm U}(1)$ iDMRG simulation.
 One pseudo first-order transition point $\phi_{f}(\chi)$ and one pseudo critical point $\phi_{c}(\chi)$ are detected as a (discontinuous) dip and a peak for each value of the bond dimension $\chi$, respectively.
 Inset: an extrapolation of the pseudo critical
points $\phi_{c}(\chi)$ is performed, as the bond dimension $\chi$ tends to infinity, assuming
$\phi_{c}(\chi)$ terminate
at the $\rm{SU}(3)$ ferromagnetic point. Thus, the data is fitted in the form:
$\phi_{c}(\chi)= 5/4 \pi+ p \chi ^q$, where  $p=-2.587$ and $q=-0.0026$.
 } \label{U1Fig1}
\end{figure}

{\it The ${\rm U}(1)$ $\rm{iDMRG}$ simulation.-}
The ${\rm U}(1)$ iDMRG algorithm with two-site
translational invariance, is exploited, targeting at a ground state, with the $z$-component of the total spin being zero,
to simulate the model (\ref{ham1}) in the same region. In Fig.~\ref{U1Fig1}, we plot the entanglement entropy $S(\phi,\chi)$
as a function of $\phi$, with the bond dimension $\chi=30,200,300,500,1000$ and $2000$.
It is found that a peak and a (discontinuous) dip appear, indicating that there are two pseudo phase transition points:
one is a pseudo critical point $\phi_c(\chi)$, consistent with the iTEBD simulation, and the other is a pseudo first-order phase transition point $\phi_{f}(\chi)$, for each value of $\chi$.  The latter separates a pseudo
critical regime with central charge $c=2$
from a pseudo fractal regime.
As an illustrative example,
for  $\chi=2000$, the pseudo first-order phase transition point $\phi_{f}(\chi)$ and the pseudo critical point $\phi_{c}(\chi)$
are located at $\phi_{f}(\chi)=1.256\pi$ and $\phi_c(\chi)=1.292\pi$, respectively.
We also perform an extrapolation of the pseudo critical
points $\phi_{c}(\chi)$, as the bond dimension $\chi$ tends to infinity, assuming that the pseudo critical
points $\phi_{c}(\chi)$ terminate at the $\rm{SU}(3)$ ferromagnetic point. Thus, the fitting function takes the form:
$\phi_{c}(\chi)= 5/4 \pi+ p \chi ^q$, where  $p=-2.587$ and $q=-0.0026$~\cite{footnote}, as shown in Fig.~\ref{U1Fig1}.

To characterize the pseudo fractal regime, we resort to the finite block-size scaling of the entanglement entropy $S(n)$.
As it turns out, it scales
logarithmically with the block size $n$~\cite{Doyon}:
\begin{equation}
 {S(n) = \frac{d_f}{2}\ln n} + b.
\label{FractalD}
\end{equation}
Here, $d_f$ is the fractal dimension, and $b$ is an additive constant. For three different values of $\phi$: $\phi=1.26\pi, 1.254\pi$ and $1.252\pi$ in the pseudo fractal regime, we perform
the best linear fit, yielding that $d_f = 1$, with a relative error being less than $4\%$,
consistent with a general statement that the fractal dimension $d_f$ is equal to the number of the (pseudo) GMs
of type-B~\cite{shiqq}(also cf. Sec.~SVII in SM).

From the finite-entanglement scaling (\ref{fes}), the central charge $c$
is extracted for $\phi=1.26\pi,1.27\pi,1.28\pi$ and $1.29\pi$ in the pseudo critical regime, yielding
$c = 2$, with a relative error being less than $4\%$. This is in agreement with that from the iTEBD simulation.
Meanwhile, the central charge $c$ is extracted from a pseudo
critical point $\phi_{c}(\chi)$ between the pseudo critical regime and the dimerized regime, which yields $c=1.833$, with a relative error being less than $8.4\%$, compared to the exact value $c=2$.

Both the pseudo fractal regime and the pseudo critical regime may be characterized in terms of the pseudo order parameter $\langle O_j \rangle$.
From the ${\rm U}(1)$ iDMRG simulation, one finds that only
 $K^j_{4}=I-3/2 (S_{j}^{z})^2$
 yields a non-zero expectation value.
Therefore, the pseudo local order parameter $\langle O_j \rangle$ may be replaced by $\langle (S_{j}^{z})^2 \rangle$.
Hence, the pseudo first-order phase transition point $\phi_f(\chi)$ and the pseudo critical points $\phi_c(\chi)$ are determined from the pseudo local order parameter $\langle (S_{j}^{z})^2 \rangle$, consistent with those
from the entanglement entropy $S(\phi,\chi)$. We stress that the two regimes may be distinguished by means of the pseudo local order parameter $\langle (S_{j}^{z})^2 \rangle$, with pseudo SSB from ${\rm SU}(2)$ to  ${\rm U}(1)$ in two distinct ways. In the pseudo fractal regime, we have $\langle (S^{z}_{j})^2 \rangle>2/3$: $\rm{SU}(2)$ is spontaneously broken to $\rm{U}(1)$, with two broken generators corresponding to one pseudo GM of type-B.
In contrast,
in the pseudo critical regime, we have $\langle (S^{z}_{j})^2 \rangle<2/3$:
$\rm{SU}(2)$ is spontaneously broken to $\rm{U}(1)$, with two broken generators corresponding to two pseudo GMs of type-A.
Note that $\langle (S^{z}_{j})^2\rangle$ approaches $2/3$ to recover the ${\rm SU}(2)$ invariance, as the bond dimension $\chi$ goes to infinity.

The presence of the pseudo first-order phase transition points $\phi_{f}(\chi)$ from the $\rm{U}(1)$ iDMRG simulation makes it possible to refine the two scenarios from the iTEBD. In particular, this entails an important implication for the second scenario. Suppose the pseudo critical points $\phi_{c}(\chi)$ terminate at a critical point $\phi_c$, as the bond dimension $\chi$ tends to infinity. Then, the critical point $\phi_c$ must be
located at a value of $\phi$ greater than $\phi_o$. Here, $\phi_o$ denotes an onset point where the pseudo fractal regime sets in, which reflects ferromagnetic fluctuations from the $\rm{SU}(3)$ ferromagnetic point. Physically, this amounts to stating that the pseudo fractal regime is not
allowed to extend to a point beyond the critical point $\phi_c$, if it exists. For the first scenario,  both $\phi_{f}(\chi)$ and $\phi_c(\chi)$
approach the $\rm{SU}(3)$ ferromagnetic point, as the bond dimension $\chi$ tends to infinity.

%

\begin{figure}
\includegraphics[width=0.37\textwidth]{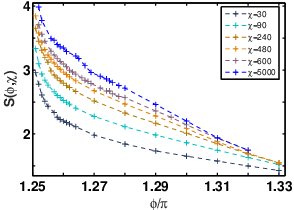}
\caption{The entanglement entropy $S(\phi,\chi)$ as a function of $\phi$,
with the bond dimension $\chi=30,90,240,480,600$ and $5000$, from the $\rm{SU}(2)$ iDMRG simulation. Note that no phase transition is detected in the region $(1.262\pi, 1.33\pi]$, in which the simulation results are reliable.
}\label{U2Fig1}
\end{figure}

{\it The $\rm{SU}(2)$ $\rm{iDMRG}$ simulation.-}
We exploit the $\rm{SU}(2)$ iDMRG algorithm to simulate the model (\ref{ham1}) in the region $(1.251\pi, 1.33\pi]$.
In Fig.\ref{U2Fig1}, we plot the entanglement entropy $S(\phi,\chi)$ as a function of $\phi$, with
the bond dimension $\chi=30,90,240,480,600$ and $5000$.  The entanglement entropy $S(\phi,\chi)$
monotonically increases, as the bond dimension $\chi$ increases.
Hence, the entanglement entropy $S(\phi,\chi)$ is smooth in the region $\phi\in(1.251\pi,1.33\pi]$, indicating
that no phase transition occurs.
This observation is confirmed from the behaviors of the dimerized order parameter $\langle D_{j,j+1} \rangle$ and the ground state fidelity $d(\phi_1,\phi_2)$ per lattice site ~\cite{zhou} ( cf.~SV in SM).

Therefore, we are led to the conclusion that the second scenario is unlike to be valid. That is, the dimerized phase is
extended up to the $\rm{SU}(3)$  ferromagnetic point, and the phase transition from the ferromagnetic phase to the dimerized phase is direct, consistent with Ref.~\cite{Fath2}.

One might speculate that the simulation results from the $\rm{SU}(2)$
iDMRG algorithm alone are sufficient to rule out the possibility for the existence of the critical nematic phase. However, this is not the case, simply because the accuracies become worse, as the $\rm{SU}(3)$ ferromagnetic point is approached.
Actually, it is the occurrence of the pseudo fractal regime in the ${\rm U}(1)$ iDMRG simulation that makes it unnecessary to get close to the $\rm{SU}(3)$ ferromagnetic point.  Indeed,  the simulation results from the $\rm{SU}(2)$ iDMRG algorithm are less reliable when $\phi$ is in the region $(1.251\pi,1.262\pi]$. However, the pseudo fractal regime in the ${\rm U}(1)$ iDMRG simulation extends at least up to $1.265\pi$. That means $\phi_o$ is at least $1.265\pi$. This in turn explains why it is sufficient to rule out the existence of a critical nematic phase, if no phase transition is detected in the region $(1.262\pi, 1.33\pi]$.

 {\it The origins of the pseudo  fractal and critical regimes in the tensor network simulations.-}
The pseudo fractal regime only occurs in the $\rm{U}(1)$ iDMRG simulation. To unveil the origin of the pseudo fractal regime, we need first to clarify the underlying physics behind the $\rm{SU}(3)$ ferromagnetic model. Indeed, the ground states are scale but not conformally invariant~\cite{shiqq}. In fact, two GMs of type-B emerge, as a consequence of SSB from $\rm{SU}(3)$ to $\rm{SU}(2) \otimes \rm{U}(1)$, implying that the fractal dimension $d_f$ is equal to 2, since the fractal dimension $d_f$ measures the number of low-lying excitations (cf. SVII in SM). Given a ground state with the $z$-component of the total spin being zero is targeted during the $\rm{U}(1)$ iDMRG simulation, one may attribute the pseudo fractal regime to a proximity effect to highly degenerate and highly entangled ground states at the $\rm{SU}(3)$ ferromagnetic point. On the other hand, the fractal dimension $d_f$ is equal to 1 in the pseudo fractal regime, since one pseudo GM of type-B is present, as a result of pseudo SSB from $\rm{SU}(2)$ to $\rm{U}(1)$. Meanwhile, the pseudo GM of type-B smoothly evolves into one of the two GMs of type-B at the $\rm{SU}(3)$ ferromagnetic point.

The remaining question concerns what acts as the origin of the pseudo critical regime.
In Ref.~\cite{BBDmodel}, the spin-1 bilinear-biquadratic model undergoing a quadratic Zeeman effect is unveiled to exhibit the KT transitions between the XY nematic phase and the dimerized phase. That is, there is a critical nematic phase with $c=1$, which is infinitesimally close to the $\rm{SU}(3)$ ferromagnetic point. Therefore, if an extra term
$D_x\sum_{j}(S_{j}^x)^2   +D_y\sum_{j}(S_{j}^y)^2+D_z\sum_{j}(S_{j}^y)^2$ is included in the Hamiltonian (\ref{ham1}), with
$D_x$, $D_y$, and $D_z$ being some coupling constants, one may expect that there are two critical nematic phases with $c=1$, which extend up to the $\rm{SU}(3)$ ferromagnetic point, due to the local constraints $(S_{j}^x)^2+(S_{j}^y)^2+(S_{j}^y)^2 =2$ (also cf. SVIII in SM).
As a consequence, two pseudo GMs of type-A emerge in the iTEBD and $\rm{U}(1)$ iDMRG simulations, as long as it is close enough to the $\rm{SU}(3)$ ferromagnetic point. In other words, it is the proximity effect to the two critical phases with $c=1$ that constitutes the origin of the pseudo critical regime with $c=2$ in the iTEBD  and $\rm{U}(1)$ iDMRG simulations. However, the two critical phases with $c=1$ do not meet each other at the $\rm{SU}(3)$ ferromagnetic point, though asymptotically close. This implies that no $\rm{SU}(2)$ WZW model with level $k=4$ survives as a limit of the pseudo critical regime.

{\it Concluding remarks.-} We have performed extensive numerical simulations of the one-dimensional spin-1 bilinear-biquadratic model in the context of the tensor network algorithms. Our results demonstrate that different types of numerical artifacts arise from the tensor network algorithms, if different symmetries are implemented. There are two types of the artifacts, attributed to a proximity effect to a highly entangled ground state, either a conformally invariant critical regime or a scale invariant fractal regime. The artifacts may be characterized in terms of the fractal dimension~\cite{Doyon} and the counting rule of the pseudo GMs~\cite{Watanabe}.

A few closing remarks are in order.  First, the suggestion made by Chubukov~\cite{Chubukov}, though it has been ruled out, is not far from being true, in the sense that the two critical nematic phases with the central charge $c=1$  are asymptotically close to the $\rm{SU}(3)$ ferromagnetic point, though no $\rm{SU}(2)$ WZW model at level $k=4$ survives as a limit of the pseudo critical regime, as the bond dimension increases. It is
the proximity effect to the two critical nematic phases that provides a mechanism responsible for the opening of an exponentially decaying small gap away from the $\rm{SU}(3)$ ferromagnetic point. Accordingly, an essential singularity results from a catastrophe point, in a similar way to the Kosterlitz-Thouless transitions~\cite{Wang}. Second, the phase transition from the ferromagnetic phase to the dimerized phase is direct, consistent with Ref.~\cite{Fath2}. The transition point, located at the $\rm{SU}(3)$ ferromagnetic point, features scale invariant ground states, with the fractal dimension being equal to 2.
Indeed, the $\rm{SU}(2)$ ferromagnetic states are smoothly embedded into the $\rm{SU}(3)$ ferromagnetic states from the ferromagnetic phase. In contrast, the essential singularity appears if the $\rm{SU}(3)$ ferromagnetic point is approached from the dimerized phase.
As a consequence, the phase transition is not of the first-order, in contrast to the claim in Ref.~\cite{Fath2}.
Third, instead of being a crossover~\cite{Lauchli}, the artifacts involve the pseudo first-order phase transition and the pseudo critical points, thus much more complicated than anticipated.

{\it Acknowledgements.-} We thank Murray Batchelor, Sam Young Cho, John Ove Fj{\ae}restad, Javier Rodr\'{i}guez-Laguna, Silvia N. Santalla, and Germ\'{a}n Sierra for enlightening discussions.
The work is supported by the National Natural Science Foundation of China (Grant No. 11805285).


\begin{thebibliography}{99}
\bibitem{footnote} The different values of $p$ and $q$  from the iTEBD and $\rm {U(1)}$ iDMRG simulations are due to the fact that
the constraints of the same value of $\chi$ imposed on the infinite MPS representations in the two cases are different.
\end{thebibliography}
\end{document}


\title{Supplementary Material for ``absence of a critical nematic phase in the vicinity of the $\rm {SU}(3)$ ferromagnetic point for the one-dimensional spin-1 bilinear-biquadratic model"}

\author{Yan-Wei Dai}
\affiliation{Centre for Modern Physics,
Chongqing University, Chongqing 400044, The People's Republic of
China}

\author{Qian-Qian Shi}
\affiliation{Centre for Modern Physics,
Chongqing University, Chongqing 400044, The People's Republic of
China}

\author{Huan-Qiang Zhou}
\affiliation{Centre for Modern Physics,
Chongqing University, Chongqing 400044, The People's Republic of
China}

\author{Ian P. McCulloch}
\affiliation{ Department of Physics, National Tsing Hua University, Hsinchu 30013, Taiwan}
\affiliation{School of Mathematics and Physics, The University of Queensland, St. Lucia, QLD 4072, Australia}
\affiliation{Centre for Modern Physics, Chongqing University, Chongqing 400044, The People's Republic of China}

\maketitle
\section*{SI. The two $\rm{SU}(3)$ symmetries}\label{SUsymmetry}
The Hamiltonian for the one-dimensional spin-1 bilinear-biquadratic model (1) possesses the $\rm{SU}(2)$ symmetry, with the generators
$\sum_{j}S_{j}^{x}$, $\sum_{j}S_{j}^{y}$ and $\sum_{j}S_{j}^{z}$, for a generic value of the parameter $\phi$. However,
its symmetry is enlarged to $\rm{SU}(3)$ at four points $\phi=\pi/2, 3\pi/2, \phi=\pi/4$, and $\phi=5\pi/4$: one is staggered at $\phi=\pi/2$ and $\phi=3\pi/2$
~\cite{Affleck0,chenxh0}, and the other is uniform at $\phi=\pi/4$ and $\phi=5\pi/4$~\cite{sutherland0}.

At $\phi=\pi/2$ and $\phi=3\pi/2$, the staggered $\rm{SU}(3)$ symmetry is realized in terms of the spin-1 operators:
$J_{\alpha}=\sum_{j}J_{\alpha}^j$
 $(\alpha= 1, 2, \cdots, 8)$, with
$J_1=1/2\sum_{j}S_j^x$, $J_2=1/2\sum_{j}S_j^{y}$, $J_3=1/2\sum_{j}S_j^z$,
 $J_4=1-3/2\sum_{j}(-1)^{j+1}(S_j^z)^2$, $J_5=1/2\sum_{j}(-1)^{j+1}({(S_j^x)}^2-{(S_j^y)}^2)$,
 $J_6=1/2\sum_{j}(-1)^{j+1}(S_j^yS_j^z+S_j^zS_j^y)$, $J_7=1/2\sum_{j}(-1)^{j+1}(S_j^zS_j^x+S_j^xS_j^z)$ and
$J_8=1/2\sum_{j}(-1)^{j+1}(S_j^xS_j^y+S_j^yS_j^x)$.

At $\phi=\pi/4$ and $\phi=5\pi/4$, the uniform $\rm{SU}(3)$ symmetry is realized in terms of the spin-1 operators:
$K_{\alpha}=\sum_{j}K_{\alpha}^j$
 $(\alpha= 1, 2, \cdots, 8)$, with
$K_1=1/2\sum_{j}S_j^x$, $K_2=1/2\sum_{j}S_j^{y}$, $K_3=1/2\sum_{j}S_j^z$,
 $K_4=1-3/2\sum_{j}(S_j^z)^2$, $K_5=1/2\sum_{j}({(S_j^x)}^2-{(S_j^y)}^2)$,
 $K_6=1/2\sum_{j}(S_j^yS_j^z+S_j^zS_j^y)$, $K_7=1/2\sum_{j}(S_j^zS_j^x+S_j^xS_j^z)$ and
$K_8=1/2\sum_{j}(S_j^xS_j^y+S_j^yS_j^x)$.

\section*{SII. The central charge $c$ in the pseudo critical regime and at the pseudo critical points}\label{Centralcharge}

The entanglement entropy is defined as
$S=-Tr\rho\ln\rho$, which may be exploited to quantify the bipartite entanglement~\cite{Bennett0}. In our case,
the density matrix $\rho=|\varphi\rangle\langle\varphi|$ is generated from the ground state wave function $|\varphi\rangle$,
in the infinite MPS representation from the tensor network simulations. For our purpose, we consider the entanglement entropy for a semi-infinite chain, which may be
rewritten as
\begin{equation}
 {S(\phi,\chi)=-\sum_{\mu}\lambda_{\mu}^2(\phi,\chi)\ln\lambda_{\mu}^2(\phi,\chi),}
 \label{entropy}
\end{equation}
where $\lambda_{\mu}^2(\phi,\chi)$ is the Schmidt decomposition coefficients.
The finite-entanglement scaling (2)~\cite{Tagliacozzo0, Pollmann0} is performed for the $S(\phi,\chi)$
to extract the central charge
$c$ in the pseudo critical regime as well as from the pseudo critical points $\phi_{c}(\chi)$.
Note that the correlation length $\xi$ is defined in terms of the ratio between the largest eigenvalue $\varepsilon_0$
and the second largest eigenvalue
$\varepsilon_1$ of the transfer matrix: $\xi=1/\ln|\varepsilon_0/\varepsilon_1|$.

\begin{figure}
\includegraphics[width=0.35\textwidth]{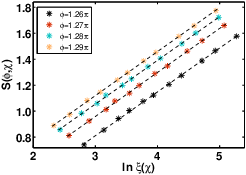}
\caption{(color online) The iTEBD simulation for the finite-entanglement scaling: $S(\phi,\chi)$ versus $\ln \xi(\chi)$, in the pseudo critical regime.
Here, $\phi=1.26\pi, 1.27\pi, 1.28\pi$ and $1.29\pi$, and the bond dimension $\chi$ ranges from $16$ to $300$.
  The central charge $c$ is extracted to be $c = 2$ in the pseudo critical regime.
 } \label{iFig4}
\end{figure}

\begin{table}
\renewcommand\arraystretch{2}
\caption{The central charge $c$ is extracted to be $c=2$ in the pseudo critical regime from the iTEBD simulation.
Here, $\phi=1.26\pi,1.27\pi,1.28\pi$ and $1.29\pi$.}
\begin{tabular}{cccccccc}
\hline\hline
      \begin{minipage}{1.2cm} $\phi$ \end{minipage}
      &\begin{minipage}{1.2cm} $1.26\pi$ \end{minipage} &
      \begin{minipage}{1.2cm} $1.27\pi$ \end{minipage} &
      \begin{minipage}{1.2cm} $1.28\pi$ \end{minipage} &
      \begin{minipage}{1.2cm} $1.29\pi$ \end{minipage} \\
\hline
 \begin{minipage}{1.2cm} $c$ \end{minipage}
 & 2.0214 & 2.0322 & 2.0184 & 2.0334 \\
 \hline\hline
\end{tabular}
\label{table1}
\end{table}
\begin{figure}
\includegraphics[width=0.35\textwidth]{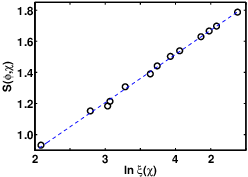}
\caption{(color online) The iTEBD simulation for the finite-entanglement scaling: $S(\phi,\chi)$ versus $\ln \xi(\chi)$, at the pseudo
critical points $\phi_c(\chi)$.
The bond dimension $\chi$ ranges from $10$ to $200$.
The central charge $c$ is extracted to be $c=1.8624$, close to the expected value $c=2$, with a relative error being less than $7\%$.
 } \label{iFig5}
\end{figure}

\subsubsection*{1. The central charge $c$ from the $\rm{iTEBD}$ simulation}
For our purpose, we focus on the region $(1.25\pi,1.5\pi]$ of the parameter $\phi$ and exploit the iTEBD to simulate the model (1). As a result, a pseudo critical point $\phi_{c}(\chi)$ is detected from the iTEBD simulation for each value of the bond dimension $\chi$. Therefore, the region is separated
into the pseudo critical regime and the dimerized regime.

In order to characterize the pseudo critical regime, the finite-entanglement scaling for the entanglement entropy $S(\phi,\chi)$ is performed for various values of $\phi$ in the pseudo critical regime, with the bond dimension $\chi$
ranging from $16$ to $300$. Here, we have randomly chosen $\phi=1.26\pi, 1.27\pi, 1.28\pi$ and $1.29\pi$, which are located in the
pseudo critical regime.
In Fig.~\ref{iFig4}, the best linear fit is exploited to estimate the central charge $c$,
which is listed in Table~\ref{table1}.
The iTEBD simulation yields that the central charge is $c=2$, with a relative error being less than $2\%$
in the pseudo critical regime.

In addition, the central charge $c$ is extracted by performing the finite-entanglement scaling from the pseudo critical points $\phi_c(\chi)$.
In Fig.~\ref{iFig5}, we plot the entanglement entropy $S(\phi,\chi)$ versus $\ln\xi(\chi)$, with the bond dimension ranging
from $10$ to $200$.
The best linear fit is exploited to estimate the central charge $c=1.8624$, with a relative error being less than $7\%$, compared to the expected value $c=2$.
Here, we stress that the less accurate central charge $c$ extracted from the pseudo critical points results from an error in determining the locations of the pseudo critical points $\phi_c(\chi)$, due to the high computational costs involved in the iTEBD simulation, which may be significantly improved if more trials are implemented.


\begin{figure}
\includegraphics[width=0.35\textwidth]{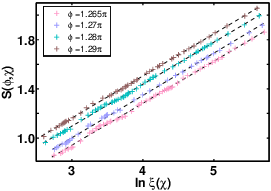}
\caption{(color online) The $\rm{U}(1)$ iDMRG simulation for the finite-entanglement scaling: $S(\phi,\chi)$ versus $\ln \xi(\chi)$,
with the bond dimension $\chi$ ranging from $30$ to $1000$. Here,
$\phi=1.265\pi, 1.27\pi, 1.28\pi$ and $1.29\pi$.
The central charge $c$ is extracted to be $c=2$ in the pseudo critical regime.
 } \label{U1Fig3}
\end{figure}

\begin{figure}
\includegraphics[width=0.35\textwidth]{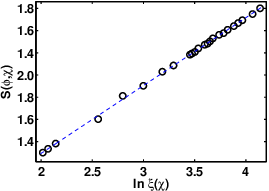}
\caption{(color online) The $\rm{U}(1)$ iDMRG simulation for the finite-entanglement scaling: $S(\phi,\chi)$
versus $\ln \xi(\chi)$, at the pseudo critical points $\phi_c(\chi)$.
The bond dimension $\chi$ ranges from $60$ to $600$.
The central charge $c$ is extracted to be $c=1.833$, with a relative error being less than $8.4\%$, compared to the expected value $c=2$.
 } \label{U1Fig4}
\end{figure}

\begin{table}
\renewcommand\arraystretch{2}
\caption{The central charge $c$ is extracted to be $c=2$ in the pseudo critical regime from the $\rm{U}(1)$ iDMRG simulation.
 Here, $\phi=1.265\pi,1.27\pi,1.28\pi$ and $1.29\pi$.}
\begin{tabular}{cccccccc}
\hline
\hline
\begin{minipage}{1.2cm} $\phi$ \end{minipage}
      &\begin{minipage}{1.2cm} $1.265\pi$ \end{minipage} &
      \begin{minipage}{1.2cm} $1.27\pi$ \end{minipage} &
      \begin{minipage}{1.2cm} $1.28\pi$ \end{minipage} &
      \begin{minipage}{1.2cm} $1.29\pi$ \end{minipage} \\
\hline
 \begin{minipage}{1.2cm} $c$ \end{minipage}
 & 2.0478 & 2.0664 & 2.0478 & 2.055 \\
 \hline\hline
\end{tabular}
\label{table2}
\end{table}

\subsubsection*{2. The central charge $c$ from the $\rm{U}(1)$ $\rm{iDMRG}$ simulation}
The $\rm{U}(1)$ iDMRG simulation is performed for the model (1) in the region $(1.25\pi,1.4\pi]$. Two pseudo phase transition points are detected from the $\rm{U}(1)$ iDMRG simulation for each value of the bond dimension $\chi$: one is a pseudo
first-order phase transition point $\phi_{f}(\chi)$, and the other is a pseudo critical point $\phi_{c}(\chi)$. The latter also occurs in the iTEBD simulation, though the former does not.
The two pseudo phase transition points separate the region into three regimes: the pseudo fractal regime, the pseudo critical regime and the dimerized regime.

The finite-entanglement scaling is performed for the entanglement entropy $S(\phi,\chi)$, with
$\phi=1.265\pi, 1.27\pi, 1.28\pi$ and $1.29\pi$, which
are located in the pseudo critical regime from the $\rm{U}(1)$ iDMRG simulation. In Fig.\ref{U1Fig3}, we plot the entanglement entropy $S(\phi,\chi)$ versus $\ln\xi(\chi)$,
with the bond dimension $\chi$ ranging from $30$ to $1000$.  The best linear fit is exploited to estimate the central charge $c$,
which is listed in Table~\ref{table2}.
The $\rm{U}(1)$ iDMRG simulation yields that the central charge is $c=2$, with a relative error being less than $4\%$ in the pseudo critical regime, consistent with the iTEBD simulation.

The central charge $c$ is also extracted by performing the finite-entanglement scaling from the pseudo critical points $\phi_c(\chi)$ between the pseudo critical regime and the dimerized regime, determined from the $\rm{U}(1)$ iDMRG simulation.
In Fig.~\ref{U1Fig4}, the best linear fit is exploited to estimate the central charge $c=1.833$, with the bond
dimension $\chi$ ranging from $60$ to $600$. As a result,  the central charge $c$ is
close to the expected value $c=2$, with a relative error being less than $8.4\%$. Similar to the iTEBD simulation,
this may be significantly improved if more trials are implemented.
%

\section*{SIII. Pseudo SSB in the pseudo critical regime: the iTEBD and the $\rm{U}(1)$ iDMRG simulations} \label{PseudoSSB}
\begin{figure}
\includegraphics[width=0.3\textwidth]{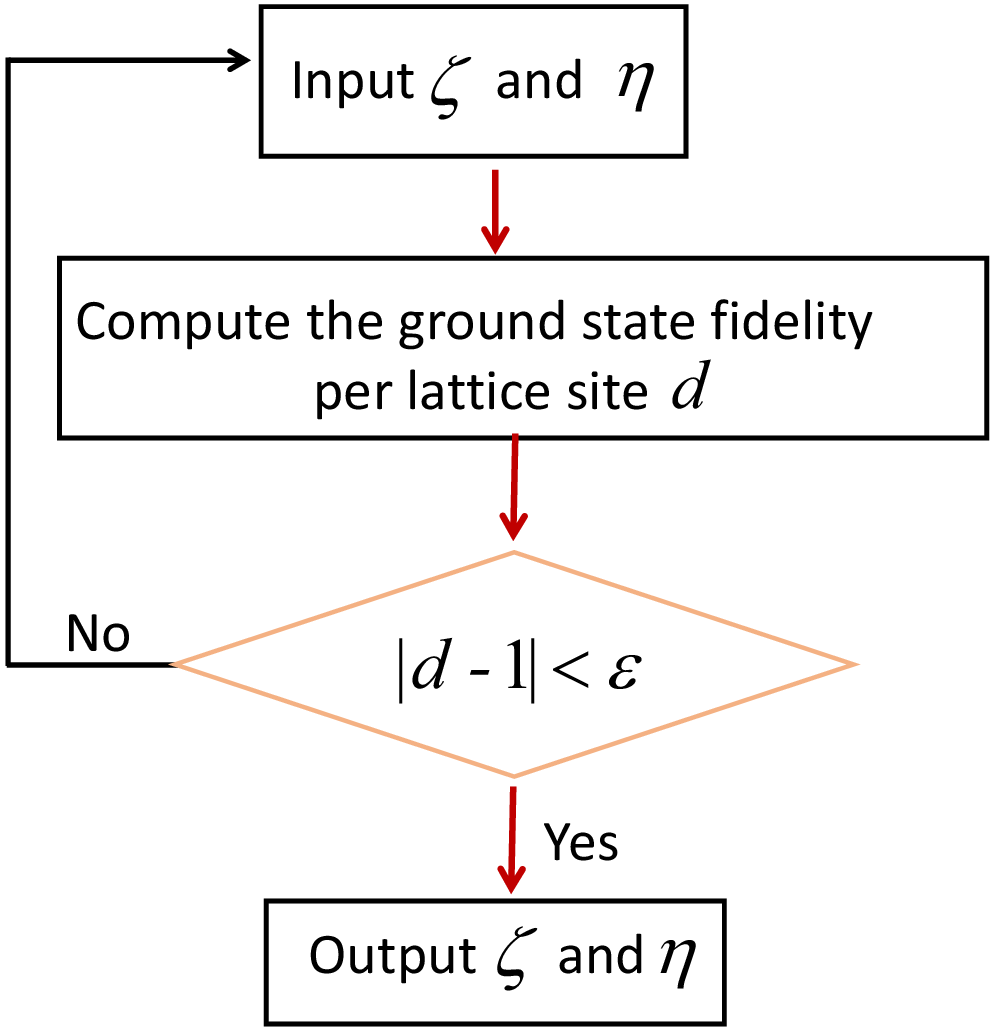}
\caption{(color online) A flow chart to determine the directional vector $\vec{n}$ for a ground state wave function
generated from the iTEBD and $\rm{U}(1)$ $\rm{iDMRG}$ simulations. Here, $\varepsilon$ denotes a preset error.
 }
\label{flowchart}
\end{figure}

To set the stage, we introduce a few notations. For a directional vector
$\vec{n}=(\cos\zeta\cos\eta,\cos\zeta\sin\eta,\sin\zeta)$, the spin component $S_j^{\vec{n}}$ along the directional vector $\vec{n}$
takes the form: $S_j^{\vec{n}}=\cos\zeta\cos\eta S_j^x+\cos\zeta\sin\eta S_j^y+\sin\zeta S_j^z$. Our task is to determine the directional vector $\vec{n}$, along which the global spin $\mathbf{S}$ is spontaneously polarized for a given ground state wave function  generated from the iTEBD and the $\rm{U}(1)$ iDMRG algorithms. For this purpose, a subroutine is developed to determine the directional vector $\vec{n}$ for spontaneous polarization as a result of pseudo SSB in the pseudo critical regime, with flow chart being shown in Fig.~\ref{flowchart}.

For a given $\zeta\in[0,\pi]$ and $\eta\in[0,2\pi]$,
we compute the ground state fidelity per lattice site $d$~\cite{zhou0} (also cf. Eq.(\ref{fidelity}) below in Sec. SV) between $|\psi\rangle$ and $U|\psi\rangle$. Here,  $|\psi\rangle$ is
 a ground state wave function generated from the iTEBD and the $\rm{U}(1)$ iDMRG algorithms, and $U={\rm exp} ({i\omega S^{\vec{n}}})$,
  with $\omega$ being a real number.
 The idea is to optimize the ground state fidelity per lattice site $d$,
when $\zeta$ and $\eta$ are varied.

The procedure is as follows.  (i) Input an initial value of $\zeta$ and $\eta$, $\zeta$ is varied  from $0$ to $\pi$, and $\eta$ is varied from $0$ to $2\pi$, respectively, with the step size being $0.01$.
(ii) Compute the ground state fidelity per lattice site $d$.
(iii) Compare the ground state fidelity per lattice site $d$ to $1$, to see if $|d-1|> \varepsilon$, with $\varepsilon$ being a preset error.
If $|d-1|> \varepsilon$, then return to (i).
If $|d-1|<\varepsilon$, then exit and save the values of $\zeta$ and $\eta$.  In our computation, $\omega$ takes an arbitrary value,
and the preset error $\varepsilon$ is set to be $\varepsilon=10^{-6}$.  Note that the final outcome should not depend on the value of $\omega$.

\subsubsection*{1. Pseudo SSB in the pseudo critical regime: the $\rm{iTEBD}$ simulation}
The iTEBD algorithm is exploited to generate ground state wave functions for a few selected values of $\phi$:
$\phi=1.26\pi,1.27\pi,1.28\pi$ and $1.29\pi$, in the pseudo critical region. Here, we choose the bond dimension $\chi$ to be $\chi=30$.
The directional vector $\vec{n}$ is determined for spontaneous polarization as a result of pseudo SSB.  The results for $\zeta$ and $\eta$
thus yielded are listed in Table~\ref{table5}.
This indicates that $\vec{n}$ is random, meaning that spontaneous polarization in any direction is possible, consistent with pseudo SSB from $\rm{SU}(2)$ to $\rm{U}(1)$.  Here, $\rm{U}(1)$ is generated from $S^{\vec{n}}$.

\begin{table}
\renewcommand\arraystretch{2}
\caption{$\zeta$ and $\eta$ for spontaneous polarization in the pseudo critical regime from the iTEBD simulation,
 with the bond dimension $\chi=30$. Here, $\phi=1.26\pi,1.27\pi,1.28\pi$ and $1.29\pi$.}
\begin{tabular}{cccccccc}
\hline
\hline
\begin{minipage}{1.2cm} $\phi$ \end{minipage}
      &\begin{minipage}{1.2cm} $1.26\pi$ \end{minipage} &
      \begin{minipage}{1.2cm} $1.27\pi$ \end{minipage} &
      \begin{minipage}{1.2cm} $1.28\pi$ \end{minipage} &
      \begin{minipage}{1.2cm} $1.29\pi$ \end{minipage} \\
\hline
 \begin{minipage}{1.2cm}$\zeta$ \end{minipage}
 & 0.8482 &  0.7540&  0.2827& 0.3142 \\
 \begin{minipage}{1.2cm} $\eta$ \end{minipage}
 & 0.7540 & 3.7071  & 2.1991 & 2.7332 \\
 \hline\hline
\end{tabular}
\label{table5}
\end{table}

\subsubsection*{2. Pseudo SSB in the pseudo critical regime: the $\rm{U}(1)$ $\rm{iDMRG}$ simulation}
The $\rm{U}(1)$ iDMRG algorithm is exploited to generate ground state wave functions for a few selected values of $\phi$: $\phi=1.268\pi, 1.27\pi, 1.28\pi$ and $1.29\pi$ in the pseudo critical regime, with the
bond dimension $\chi=40$.

The directional vector $\vec{n}$ is determined for spontaneous polarization as a result of pseudo SSB.  The results for $\zeta$ and $\eta$
thus yielded are listed in Table~\ref{table6}, meaning that $\vec{n}$ is always along the $z$ axis, as anticipated. This is
consistent with pseudo SSB from $\rm{SU}(2)$ to $\rm{U}(1)$. Here, $\rm{U}(1)$ is generated from $S^z$.

\begin{table}
\renewcommand\arraystretch{2}
\caption{$\zeta$ and $\eta$ for spontaneous polarization in the pseudo critical regime from the $\rm{U}(1)$ iDMRG
simulation, with the bond dimension $\chi=40$. Here, $\phi=1.268\pi,1.27\pi,1.28\pi$ and $1.29\pi$.}
\begin{tabular}{cccccccc}
\hline
\hline
\begin{minipage}{1.2cm} $\phi$ \end{minipage}
      &\begin{minipage}{1.2cm} $1.268\pi$ \end{minipage} &
      \begin{minipage}{1.2cm} $1.27\pi$ \end{minipage} &
      \begin{minipage}{1.2cm} $1.28\pi$ \end{minipage} &
      \begin{minipage}{1.2cm} $1.29\pi$ \end{minipage} \\
\hline
 \begin{minipage}{1.2cm}$\zeta$ \end{minipage}
 & 4.7124 & 4.7124 & 4.7124 & 4.7124 \\
 \begin{minipage}{1.2cm} $\eta$ \end{minipage}
 & 0 & 0 & 0 & 0 \\
 \hline\hline
\end{tabular}
\label{table6}
\end{table}

%
\begin{figure}
\includegraphics[width=0.37\textwidth]{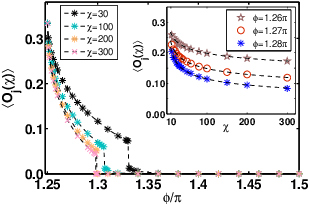}
\caption{(color online) Main: the iTEBD simulation results for the pseudo local order parameter $\langle O_j(\chi) \rangle$
as a function of $\phi$, with the bond dimension $\chi = 30, 100, 200$ and $300$. The same pseudo critical points $\phi_c(\chi)$ are detected
as those from the entanglement entropy $S(\phi,\chi)$.
Inset: the pseudo local order parameter $\langle O_j(\chi) \rangle$ may be scaled down to zero, when the bond dimension $\chi$ tends to infinity, in the pseudo critical phase. Here, the fitting function is  $\langle O_j(\chi) \rangle=s \chi^ t$, with $s=0.3528, 0.3704$ and $0.3970$ and  $t=-0.1249, -0.1985$ and $ -0.2701$, for $\phi=1.26\pi, 1.27\pi$ and $1.28\pi$, respectively.
 } \label{iFig6}
\end{figure}

\subsection*{SIV. The pseudo local order parameter $\langle O_j(\chi) \rangle$ and the dimerized local order parameter
$\langle D_{j,j+1}(\chi)\rangle$}\label{OrderDimer}
As noted in Ref.~\cite{Wang0}, the tensor network algorithms in the infinite MPS representation naturally lead to infinitely degenerate ground states in a critical regime, due to the finiteness of the
bond dimension. This results in the so-called pseudo continuous SSB~\cite{Wang0},
with the pseudo symmetry-breaking order being quantified in terms of a pseudo local order parameter $\langle O_j(\chi) \rangle$. Here, we note that the pseudo GMs involved are of type-A, if we adapt the classification of the GMs in effective field theories~\cite{Watanabe0} to the pseudo GMs arising from the finiteness of the bond dimension in the tensor network simulations.
In order to keep consistency with the Mermin-Wagner theorem~\cite{Mermin0},
$\langle O_j(\chi) \rangle$
must be scaled down to zero, when the bond dimension $\chi$ tends to infinity.
Here, by the Mermin-Wagner theorem we mean a statement that
no continuous symmetry group is spontaneously broken in one dimensional quantum many-body systems, if the GMs involved are of type-A. As a consequence, the Mermin-Wagner theorem does not rule out the possibility for continuous SSB in one dimensional many-body quantum systems, if the GMs involved are of type-B~\cite{Watanabe0,shiqq0}.

Physically, the finiteness of the bond dimension
$\chi$ in the implementation of the tensor network algorithms in the infinite MPS representation amounts to introducing extra long-range
interactions into the model Hamiltonian, which turns a quantum many-body system in one spatial dimension into one
in higher than one spatial dimensions, if one insists to consider only the short-range interactions. This makes it possible for continuous SSB with the pseudo GMs of type-A to occur in numerical simulations of one dimensional quantum many-body systems, which constitutes the origin of a pseudo critical regime.

For the model (1), a pseudo critical regime with the central charge $c=2$ does occur, as a result of pseudo SSB from $\rm{SU}(2)$ to
$\rm{U}(1)$, which has been confirmed in Sec. SIII. Given the directional vector $\vec{n}$ for spontaneous polarization is random, the three spin components $S^x$, $S^y$, and $S^z$ are broken, but leaving $S^{\vec{n}}$ invariant. That is, $\langle S^x_j \rangle=0$,
$\langle S^y_j \rangle=0$, and $\langle S^z_j \rangle=0$ in the iTEBD simulation.  As it turns out, the remaining five local components of the eight generators, $K^j_{\alpha}$ ($\alpha=1,...,8$), of the $\rm{SU}(3)$ group, discussed in Sec. SI, yield non-zero expectation values, thus constituting a vector which acts as a pseudo local order parameter. However, it is convenient to choose a pseudo local order parameter, which does not depend on an initial state. For this purpose,
we adopt $\langle O_j(\chi) \rangle$, defined in (3), as the pseudo local order parameter. Note that $\langle K^j_1 \rangle$,  $\langle K^j_2 \rangle$, and  $\langle K^j_3 \rangle$, which are essentially $\langle S^x_j \rangle$,
$\langle S^y_j \rangle$, and $\langle S^z_j \rangle$, are included in (3), since they are all zero.
Indeed, the (Cartan) metric tensor $g_{\alpha\beta}$ is defined as $g_{\alpha\beta}=\sum_{\delta,\epsilon=1}^8\gamma_{\alpha\delta\epsilon}\gamma_{\beta\delta\epsilon}$,
with $\gamma_{\alpha\beta\epsilon}$ being the structural constants of the $\rm{SU}(3)$ group: $[K_\alpha,K_\beta]=i\sum_{\delta=1}^8\gamma_{\alpha\beta\delta}K_\delta$. Alternatively,  the metric tensor $g_{\alpha\beta}$ may be defined through the Killing form as $g_{\alpha\beta}=B(K_\alpha,K_\beta)$~\cite{Killing0}. The independence of $\langle O_j(\chi) \rangle$ on an initial state is confirmed.

In addition, the dimerized local order parameter $\langle D_{j,j+1}\rangle$ is introduced to characterize the dimerized phase.
The dimerized local order parameter $\langle D_{j,j+1}(\chi)\rangle$, together with the pseudo local order parameter $\langle O_j(\chi) \rangle$, provide other means to confirm the pseudo phase transition points detected from the entanglement entropy $S(\phi,\chi)$.

\begin{figure}
\includegraphics[width=0.35\textwidth]{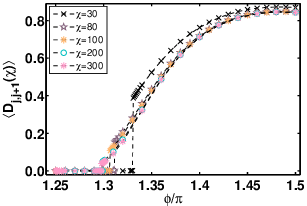}
\caption{(color online) The iTEBD simulation results for the dimerized local order parameter $\langle D_{j,j+1}(\chi) \rangle $
as a function of $\phi$, with the bond dimension $\chi = 30,  80, 100, 200$ and $300$. The same pseudo critical points $\phi_c(\chi)$
are detected
as those from the entanglement entropy $S(\phi,\chi)$.
 } \label{iFig7}
\end{figure}

 \subsubsection*{1. The pseudo local order parameter $\langle O_j(\chi) \rangle$ and the dimerized local order parameter $\langle D_{j,j+1}(\chi)\rangle$: the $\rm{iTEBD}$ simulation}

 In Fig.~\ref{iFig6}, we plot the pseudo local order parameter $\langle O_j(\chi) \rangle$
 as a function of $\phi$.  From this we see the same pseudo critical points $\phi_c(\chi)$ as those detected from the entanglement entropy $S(\phi,\chi)$.
 Specifically, the pseudo critical points $\phi_c(\chi)$ between the pseudo critical regime and the dimerized regime are located at
$\phi_{c}(\chi)=1.33\pi,1.306\pi,1.299\pi$ and $1.298\pi$ for the bond dimension $\chi=30,100,200$ and $300$, respectively.

 We remark that the pseudo local order parameter $\langle O_j(\chi) \rangle$ is nonzero in the pseudo critical phase, due to the finiteness of the bond dimension $\chi$, which is scaled down to zero,  as the bond dimension $\chi$
tends to infinity. This is shown in Fig.~\ref{iFig6}, as required to keep consistency
with the Mermin-Wagner theorem~\cite{Mermin0}.

We plot the dimerized local order parameter $\langle D_{j,j+1}(\chi)\rangle$ in Fig.~\ref{iFig7},
as a function of $\phi$ in the region $(1.25\pi, 1.5\pi]$,
with the bond dimension $\chi=30,80,100,200$ and $300$. Note that the dimerized local order parameter $\langle D_{j,j+1}(\chi)\rangle$ tends to be saturated, as the
bond dimension $\chi$ increases.
The same pseudo critical points $\phi_c(\chi)$ are detected as those from
the pseudo local order parameter $\langle O_j(\chi) \rangle$ and the entanglement entropy
$S(\phi,\chi)$.

\begin{figure}
\includegraphics[width=0.35\textwidth]{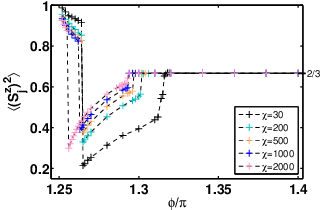}
\caption{(color online) The $\rm{U}(1)$ iDMRG simulation results for the pseudo local order parameter $\langle (S_j^z)^2 \rangle$
as a function of $\phi$, with the bond dimension $\chi = 30,  200, 500, 1000$ and $2000$. The same pseudo first-order transition points
$\phi_f(\chi)$ and pseudo critical points $\phi_c(\chi)$ are detected
as those from the entanglement entropy $S(\phi,\chi)$.
 } \label{U1Fig5}
\end{figure}

\begin{figure}
\includegraphics[width=0.35\textwidth]{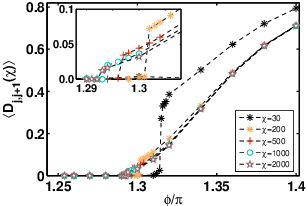}
\caption{(color online) The $\rm{U}(1)$ iDMRG simulation results for the dimerized local order parameter
 $\langle D_{j,j+1}(\chi) \rangle$ as a function of $\phi$.  The inset
shows a magnification of the dimerized
order parameter in the region $(1.288\pi, 1.305\pi)$, with the bond dimension $\chi=200,500,1000$ and $2000$. The same pseudo critical points $\phi_c(\chi)$ are detected
as those from the entanglement entropy $S(\phi,\chi)$.
} \label{U1Fig6}
\end{figure}

\subsubsection*{2. The pseudo local order parameter $\langle O_j(\chi) \rangle$ and the dimerized local order parameter $\langle D_{j,j+1}(\chi)\rangle$: the $\rm{U}(1)$ $\rm{iDMRG}$ simulation}

In the $\rm{U}(1)$ $\rm{iDMRG}$ simulation, both the pseudo fractal regime and the pseudo critical regime may be characterized in terms of
the pseudo order parameter $\langle O_j(\chi) \rangle$. Note that, since the $z$-component of the total spin is preserved to be zero,
only $K^j_{4}=I-3/2 (S_{j}^{z})^2$ yields a non-zero expectation value.
Therefore, the pseudo local order parameter $\langle O_j(\chi) \rangle$, as defined in (3), is reduced to
$
 {\langle O_j \rangle=3-9/2\langle (S_{j}^{z})^2 \rangle}.
$
Therefore, $\langle O_j(\chi) \rangle$ may be replaced by the pseudo local order parameter $\langle (S_{j}^{z})^2 \rangle$.
 In Fig.\ref{U1Fig5}, we plot the pseudo local order parameter $\langle (S_{j}^{z})^2 \rangle$
as a function of $\phi$, for the bond dimension
 $\chi=30, 200, 500, 1000$ and $2000$, respectively.
From this we observe that two pseudo phase transition points are detected from the pseudo local order parameter $\langle (S_{j}^{z})^2 \rangle$, consistent with those
from the entanglement entropy $S(\phi,\chi)$. Specifically,
the pseudo first-order phase transition points $\phi_{f}(\chi)$ and the pseudo critical points $\phi_{c}(\chi)$ are located at
$\phi_{f}(\chi)=1.265\pi,1.265\pi, 1.265\pi,1.263\pi,1.256\pi$ and
$\phi_{c}(\chi)=1.315\pi,1.301\pi,1.296\pi,1.293\pi,1.292\pi$, with the bond dimension $\chi=30,200,500,1000$ and $2000$,
respectively.
In the pseudo fractal regime, we have  $\langle (S^{z}_{j})^2 \rangle>2/3$. Therefore, $\rm{SU}(2)$ is spontaneously broken
to $\rm{U}(1)$, with two broken generators yielding one pseudo GM of type-B. This implies that the fractal dimension $d_f $, which measures the number of the pseudo GMs of type-B~\cite{shiqq0}, is equal to 1.
In contrast,
in the pseudo critical regime, we have $\langle (S^{z}_{j})^2 \rangle<2/3$. Therefore,
$\rm{SU}(2)$ is spontaneously broken to $\rm{U}(1)$, with two broken generators yielding
two pseudo GMs of type-A. This implies that the central charge $c$, which measures the number of gapless excitations, equal to 2,
since the two pseudo GMs of type-A evolves into two gapless excitations, as the bond dimension $\chi$ tends to infinity.  Note that $\langle (S_j^z)^2\rangle$ approaches $2/3$ to recover the $\rm{SU}(2)$ symmetry, as the bond dimension $\chi$ increases.

In addition,  the dimerized local order parameter is plotted in Fig.~\ref{U1Fig6} as a function of $\phi$, with $\chi=200,500,1000$ and $2000$, from the $\rm{U}(1)$ iDMRG simulation.  The inset shows a magnification of the dimerized local
order parameter $\langle D_{j,j+1}(\chi)\rangle$ in the region $(1.288\pi, 1.305\pi)$, with $\chi=200,500,1000$ and $2000$.
As a result, the pseudo critical points $\phi_{c}(\chi)$ between the dimerized regime and the pseudo critical regime are detected,
consistent with those from the pseudo local order parameter $\langle (S^{z}_{j})^2 \rangle$ as well as the entanglement entropy $S(\phi,\chi)$.

\begin{figure}
\includegraphics[width=0.35\textwidth]{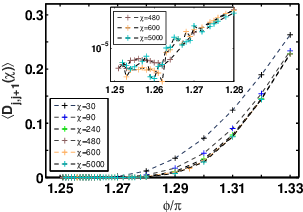}
 \caption{The $\rm{SU}(2)$ iDMRG simulation results for the dimerized local order parameter
  $\langle D_{j,j+1}(\chi) \rangle$ as a function of $\phi$.  The inset
shows a magnification of the dimerized
order parameter in the region $(1.251\pi, 1.28\pi]$, with the bond dimension $\chi=480,600$ and $5000$.
     }\label{U2Fig2}
\end{figure}

\subsubsection*{3. The dimerized local order parameter $\langle D_{j,j+1}(\chi)\rangle$ from the $\rm{SU}(2)$ $\rm{iDMRG}$ simulation}

 The $\rm{SU}(2)$ iDMRG simulation is performed for the model (1) in the region
 $(1.251\pi,1.33\pi]$. In Fig.\ref{U2Fig2}, we plot the dimerized local order parameters $\langle D_{j,j+1}(\chi)\rangle$
 as a function of $\phi$,  with the bond dimension $\chi$ ranging from $30$ to $5000$.
The inset shows a magnification of the dimerized local order parameter $\langle D_{j,j+1}(\chi)\rangle$ as a function of $\phi$,
with the bond dimension $\chi=480,600$ and $5000$, in the region $(1.251\pi, 1.28\pi]$.
Fig.~\ref{U2Fig2} indicates that the dimerized order parameter $\langle D_{j,j+1}(\chi)\rangle$
 decreases, as $\phi =5\pi/4$ is approached. However, we remark that the accuracies get worse, and so the simulation results
 become less reliable, when $\phi$ gets close to $5\pi/4$.
\section*{SV. Ground state fidelity per lattice site $d(\phi_1,\phi_2)$ from the $\rm{SU}(2)$ iDMRG simulation}\label{gsFidelity}

For two given ground states $|\psi({\phi_1})\rangle$ and $|\psi ({\phi_2}) \rangle$, the ground state fidelity $F=\langle\psi (\phi_{1})|\psi (\phi_{2})\rangle|$ asymptotically scales as
$F(\phi_{1},\phi_{2})\sim d(\phi_{1},\phi_{2})^L$,
with $L$ being the system size. Here, $d(\phi_1,\phi_2)$ is the ground state fidelity per lattice site, which is well-defined in
the thermodynamic limit~\cite{zhou0}:
 \begin{equation}
 {\ln d(\phi_{1},\phi_{2})\equiv \lim_{L\rightarrow\infty}\frac{\ln F(\phi_{1},\phi_{2})}{L}}.
 \label{fidelity}
\end{equation}
As a convention, we choose $|\psi (\phi_{1})\rangle$ as a reference state, and regard $d(\phi_{1},\phi_{2})$ as a function of $\phi_{2}$ for fixed $\phi_{1}$.

\begin{figure}
 \includegraphics[width=0.35\textwidth]{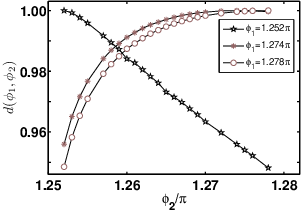}
     \caption{The $\rm{SU}(2)$ iDMRG simulation results for the ground state fidelity per lattice site $d(\phi_1,\phi_2)$ as a function of $\phi_2$,
      with a reference state $|\psi (\phi_{1})\rangle$ located at
      $\phi_1=1.252\pi, 1.274\pi$ and $1.278\pi$.
      Here, the bond dimension $\chi$ is $5000$.
     }\label{U2Fig3}
\end{figure}

In Fig.\ref{U2Fig3}, we plot the ground state fidelity per lattice site $d(\phi_1,\phi_2)$
as a function of $\phi_2$,  with $\chi=5000$, in the region $(1.251\pi, 1.28\pi)$.
Here, the reference state $|\psi (\phi_{1})\rangle$  has been chosen to be located at $\phi_1=1.252\pi, 1.274\pi$ and $1.278\pi$, respectively.
No pinch point is observed, indicating that there is no phase transition in the region $(1.251\pi, 1.28\pi)$.
%

\section*{SVI. Two scenarios arising from the tensor network simulations}\label{Twoscenarios}

For each value of the bond dimension $\chi$,  the iTEBD simulation yields a pseudo critical point $\phi_c(\chi)$, which may be
detected by means of the entanglement entropy $S(\phi,\chi)$, the pseudo local order parameter $\langle O_j(\chi)\rangle$ and the dimerized local order parameter $\langle D_{j,j+1}(\chi)\rangle$, as shown in Fig.~{2}, Fig.~\ref{iFig6} and Fig.~\ref{iFig7}, respectively.
The central charge $c$ is estimated to be $c=2$, with a reasonable relative error, in the pseudo critical regime.
One may perform an extrapolation of the pseudo critical point $\phi_c(\chi)$ to see if they terminate at the $\rm{SU}(3)$ ferromagnetic point.
In Fig.~{2}, we have shown that such an extrapolation is possible if one requires the pseudo critical points $\phi_c(\chi)$ to terminate at the $\rm{SU}(3)$ ferromagnetic point. However, the same extrapolation works well if the pseudo critical points $\phi_c(\chi)$  terminate somewhere next to the $\rm{SU}(3)$ ferromagnetic point, within the accuracies achievable.
Therefore, two scenarios arise: (i) the pseudo critical points $\phi_c(\chi)$ terminate at the $\rm{SU}(3)$ ferromagnetic point when the bond dimension $\chi$ tends to infinity;
(ii) the pseudo critical points $\phi_c(\chi)$ terminate at a point $\phi_c$ away from the $\rm{SU}(3)$ ferromagnetic point.  Here,  $\phi_c$ is defined to be a point, to which $\phi_c(\chi)$ approaches, when the bond dimension $\chi$ tends to infinity.

\begin{figure}
\includegraphics[width=0.48\textwidth]{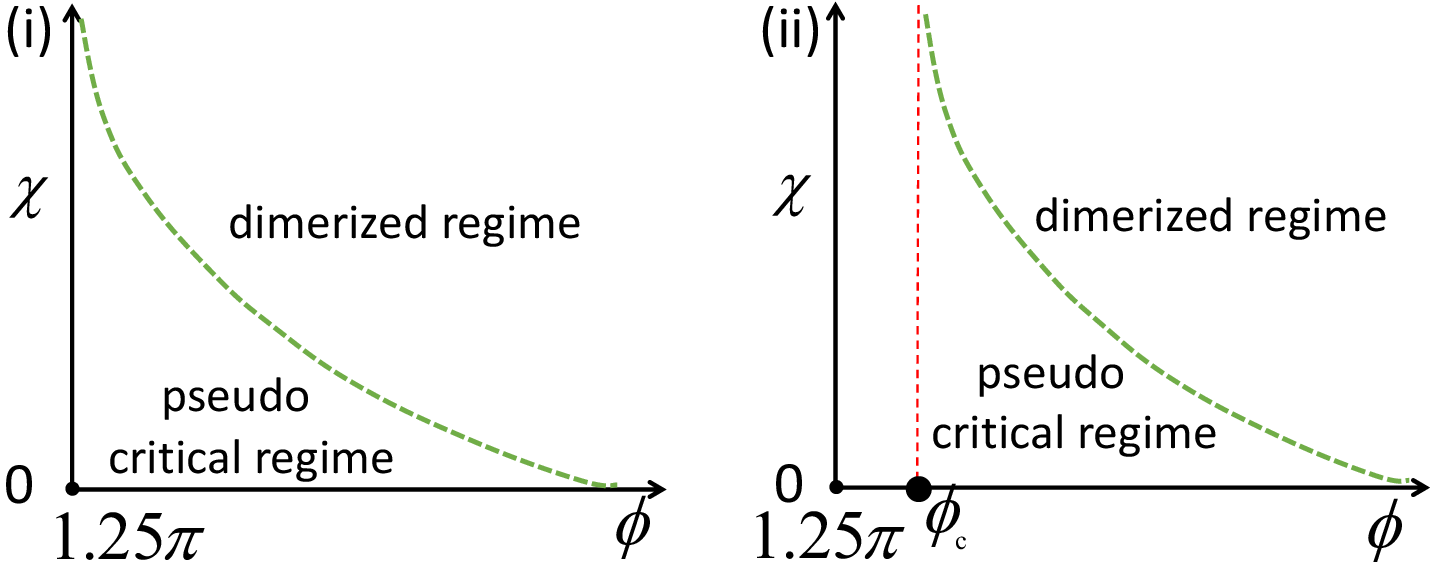}
\caption{(color online) A sketch of the two scenarios from the iTEBD simulation for the spin-1
bilinear-biquadratic model in the vicinity of the $\rm{SU}(3)$ ferromagnetic point.
 }
\label{scenario1}
\end{figure}

The two scenarios from the iTEBD simulation have been sketched in Fig.\ref{scenario1}. If the first scenario is valid, then the phase transition from the ferromagnetic phase to the dimerized phase is direct, with the phase transition point located at the $\rm{SU}(3)$ ferromagnetic point.
 If the second scenario is valid, then there are two phase transitions from the ferromagnetic phase to the dimerized phase: one is located at
 the $\rm{SU}(3)$ ferromagnetic point, and the other is located at the point $\phi_c$. Hence, there is a critical nematic phase between the ferromagnetic phase and the dimerized phase.

In the $\rm{U}(1)$ iDMRG simulation, we target at a ground state with the $z$-component of the total spin being zero. As shown in Fig.~{3}, Fig.~\ref{U1Fig5} and Fig.~\ref{U1Fig6}, from the entanglement entropy $S(\phi,\chi)$, the pseudo local order parameter $\langle O_j(\chi)\rangle$ and
the dimerized local order parameter $\langle D_{j,j+1}(\chi)\rangle$,  two phase transitions are detected: one is the pseudo first-order phase
transition points $\phi_f(\chi)$, and the other is the pseudo critical points $\phi_c(\chi)$, which have been captured in the iTEBD simulation. The occurrence of the pseudo first-order phase transition points $\phi_f(\chi)$ makes it possible to refine the two scenarios.

\begin{figure}
\includegraphics[width=0.48\textwidth]{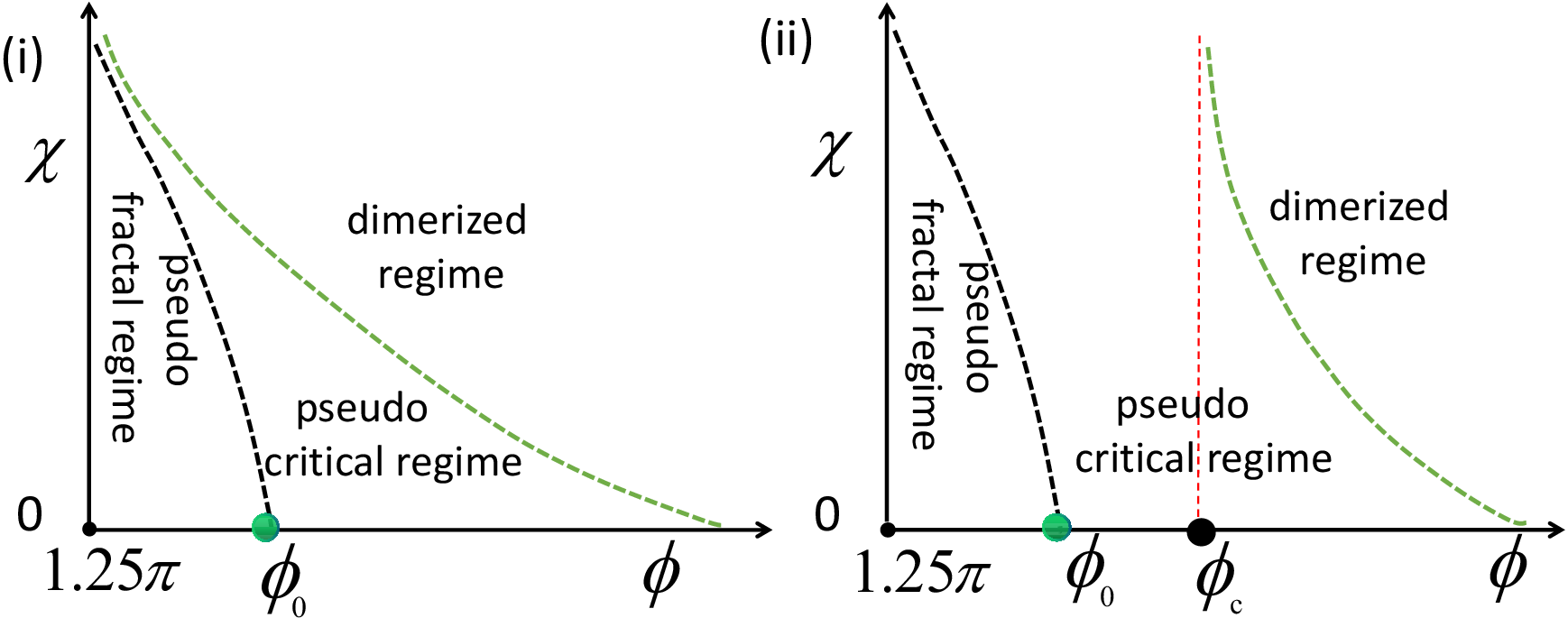}
\caption{(color online)  A sketch of the two scenarios from the $\rm{U}(1)$ iDMRG simulation for the spin-1
bilinear-biquadratic model in the vicinity of the $\rm{SU}(3)$ ferromagnetic point.
}
\label{scenario2}
\end{figure}

For the first scenario, the situation is simple: the pseudo critical regime from the iTEBD simulation is now divided into a pseudo fractal regime and a pseudo critical regime, with the pseudo first-order phase transition separating the two regimes. For the second scenario, one may argue that the pseudo fractal regime should not extend up to a point beyond $\phi_c$. That is, $\phi_c$ must always be in the pseudo critical regime for any value of the bond dimension $\chi$. Actually, performing a truncation by means of the bond dimension $\chi$ amounts to introducing an energy scale so that a gapped ground state becomes gapless, as long as the gap is small enough, compared to the energy scale thus introduced. With this in mind, we may expect that a pseudo critical point appears as a shift of the critical point $\phi_c$, but both are always in the pseudo critical regime for any value of the bond dimension $\chi$. In addition, the pseudo fractal regime reflects fluctuations from the $\rm{SU}(3)$ ferromagnetic point. Hence, it has to terminate at a point $\phi_o$, which represents the onset of the ferromagnetic fluctuations from  the $\rm{SU}(3)$ ferromagnetic point. Therefore, we have  $\phi_o < \phi_c$. This is in accord with our $\rm{U}(1)$ iDMRG simulation results.
In practice, $\phi_o$ may be replaced by the pseudo first-order phase transition point $\phi_f(\chi)$ corresponding to the smallest value of the bond dimension $\chi$ achievable.
The two scenarios from the $\rm{U}(1)$ iDMRG simulation are sketched in Fig.~\ref{scenario2}.

We stress that the pseudo first-order phase transition points $\phi_f(\chi)$ must terminate at the $\rm{SU}(3)$ ferromagnetic point, when the bond dimension $\chi$ goes to infinity. This is due to the fact that the pseudo fractal regime, characterized in terms of the fractal dimension $d_f=1$ (cf. Sec.~SVII), results from the proximity effect to the ${\rm SU(3)}$ ferromagnetic point. Note that the pseudo first-order phase transition points $\phi_f(\chi)$ are quite robust, when the bond dimension $\chi$ increases,
as seen from Fig.~{3} and Fig.~\ref{U1Fig5}.

However, the second scenario, sketched in Fig.~\ref{scenario2}, is ruled out from the $\rm{SU}(2)$ iDMRG simulation, given no phase transition is detected in the region $\phi_o<\phi<\phi_c$, as shown in Fig.~{4}, Fig.\ref{U2Fig2} and Fig.\ref{U2Fig3}, by means of the entanglement entropy $S(\phi,\chi)$, the dimerized local order parameter $\langle D_{j,j+1}(\chi)\rangle$ and the ground state fidelity per lattice site $d$, respectively. Here, $\phi_o$ is approximated by the pseudo first-order phase transition point $\phi_f(\chi)=1.265\pi$ corresponding to the bond dimension $\chi=30$, and $\phi_c$ is approximated by the pseudo critical point $\phi_c(\chi)=1.292\pi$ corresponding to the bond dimension $\chi=2000$. We remark that the region $1.262 \pi<\phi<1.33\pi$ well covers $(\phi_o, \phi_c)$.
As shown in Fig.~\ref{U2Fig2}, the dimerized order parameter $\langle D_{j,j+1}(\chi)\rangle$ at the point $\phi_o=1.265\pi$ from the $\rm{SU}(2)$ iDMRG simulation is in the order of magnitude $10^{-5}$, indicating that the $\rm{SU}(2)$ iDMRG simulation results are reliable.

Therefore, we conclude that the phase transition from the ferromagnetic phase to the dimerized phase is direct, thus ruling out the existence of a critical nematic phase in the vicinity of the $\rm{SU}(3)$ ferromagnetic point.

%

\begin{figure}
\includegraphics[width=0.32\textwidth]{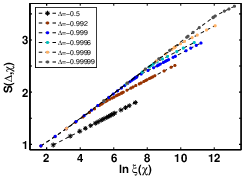}
\caption{(color online) The $\rm{U}(1)$ iDMRG simulation for the finite-entanglement scaling: $S(\Delta,\chi)$ versus $\ln \xi(\chi)$,
for the spin-$1/2$ XXZ model for $\Delta=-0.5,-0.992,-0.999,-0.9996,-0.9999$ and $-0.99999$ in the vicinity of the $\rm{SU}(2)$ ferromagnetic point
$\Delta=-1$.
 }
\label{XXZS}
\end{figure}



\section*{SVII. The fractal dimension $d_f$ in the pseudo fractal regime: the spin-$1/2$ XXZ model and the spin-1 bilinear-biquadratic model}\label{FractalDf}

According to the finite block-size scaling (4) of the entanglement entropy $S(n)$, we are able to extract the fractal dimension
$d_f$ from the tensor network simulations for the spin-1 bilinear-biquadratic model in the pseudo fractal regime. We emphasize that such a fractal regime only occurs in the $\rm{U}(1)$ iDMRG simulation.

For an illustrative purpose, we include our simulation results for the spin-$1/2$ quantum XXZ model.
We remark that it is necessary to develop a subroutine to efficiently calculate the block entanglement entropy $S(n)$.  As it turns out,
the fractal dimension $d_f$ is equal to the number of the pseudo GMs of type-B, which may be regarded as an adaptation of a general statement~\cite{shiqq0} about the connection between the fractal dimension and the number of the GMs of type-B to the numerical artifacts.

\subsubsection*{1. The fractal dimension  $d_f$ for the spin-$1/2$ XXZ model in the vicinity of the ferromagnetic point at $\Delta=-1$}
 The Hamiltonian for the spin-$1/2$ anisotropic XXZ model takes the form:
\begin{equation}
 H_{\rm{XXZ}} = \sum_{j}
 \left[S^x_jS^x_{j+1}+S^y_jS^y_{j+1}+\Delta S^z_jS^z_{j+1}
       \right],
 \label{hxxz}
\end{equation}
where $\Delta$ denotes the anisotropic coupling parameter.  At $\Delta=-1$,
the model possesses the $\rm{SU}(2)$ symmetry, generated by
$\sum_{j}(-1)^jS_{j}^{x}$, $\sum_{j}(-1)^jS_{j}^{y}$ and $\sum_{j}S_{j}^{z}$, which are staggered.

The ground state phase diagram for the spin-$1/2$ XXZ model involves three distinct phases in the entire parameter region:
the ferromagnetic phase $(\Delta<-1)$,  the critical XY phase $(-1<\Delta\leq1)$ with the central charge $c=1$,
and the N\'{e}el phase $(\Delta>1)$.
In the N\'{e}el phase, the ground state wave functions are two-fold degenerate, due to SSB of the one-site translational invariance.
In the ferromagnetic phase, the ground state wave functions are two-fold degenerate and fully
polarized along the $z$ axis.
The model is special at the point $\Delta=-1$.  In fact, it possesses highly degenerate ground states
and is not conformally invariant. A proper description requires to introduce the fractal dimension $d_f$~\cite{Doyon0}. Indeed,
this description is also necessary to understand the simulation results from the $\rm{U}(1)$ iDMRG algorithm, if one targets at a ground state with the $z$-component of the total spin being zero.

In Fig.~\ref{XXZS}, the finite-entanglement scaling of the entanglement entropy $S(\chi)$ is performed for a few chosen values of $\Delta$.
An interesting observation is that, if $\Delta$ is far away from the ferromagnetic point at $\Delta=-1$, then all the data falls on a
straight line, as the correlation length $\xi(\chi)$ increases with $\chi$. If $\Delta$ gets close enough to $\Delta =-1$, the data starts
to fall on two segments with a different slope asymptotically (see also Ref.~\cite{chen0}).  Physically, this implies a crossover
from the pseudo fractal regime to the pseudo critical regime in the vicinity of $\Delta=-1$, as the bond dimension $\chi$ increases.

In the critical XY regime, the central charge $c$ may be extracted from the finite block-size scaling $S(n)\sim c/3\ln n$~\cite{cft2}, which is listed in Table~\ref{table3-1}. In the fractal regime, the fractal dimension $d_f$ may be extracted from the finite block-size scaling of the entanglement entropy $S(n)$ (4), which is listed in Table~\ref{table3-2}.

A conclusion one may draw from the simulation results is that, as $\Delta=-1$ is approached, the fractal dimension $d_f$ tends to 1, as anticipated. We remark that the $\rm{U}(1)$ iDMRG algorithm works well even when $\Delta$ is so close to $-1$, given that it still captures the crossover behavior from the pseudo fractal regime to the pseudo critical regime, as the bond dimension $\chi$ increases.

\begin{table}
\renewcommand\arraystretch{2}
\caption{The central charge $c$ in the critical XY regime is extracted from the $\rm{U}(1)$ iDMRG algorithm
 for the spin-1/2 XXZ model, with the block size $n$ being from $6$ to $24$. Here, $\Delta=-0.2$ and $-0.5$.}
\begin{tabular}{c|ccccccc}
\hline\hline
      \begin{minipage}{1.3cm}   \end{minipage}&
      \begin{minipage}{0.9cm} $\chi$ \end{minipage}
      &\begin{minipage}{1.2cm}$20$ \end{minipage} &
      \begin{minipage}{1.2cm} $40$ \end{minipage} &
      \begin{minipage}{1.2cm} $50$ \end{minipage} &
      \begin{minipage}{1.2cm} $60$ \end{minipage} \\
\hline
 \begin{minipage}{1.5cm}$\Delta=-0.2$ \end{minipage}
 &c& 1.0143 & 1.0032 & 1.0026 & 1.002  \\
 \begin{minipage}{1.5cm}$\Delta=-0.5$ \end{minipage}
 &c &1.0395 & 1.0086 & 1.0056 & 1.0038 \\
 \hline\hline
\end{tabular}
\label{table3-1}
\end{table}

\begin{table}
\renewcommand\arraystretch{2}
\caption{The fractal dimension $d_f$ in the pseudo fractal regime is extracted from the $\rm{U}(1)$ iDMRG algorithm
for the spin-1/2 XXZ model, with the block size $n$ being from $6$ to $24$. When $\Delta = -1$ is approached, $d_f$ tends to 1.}
\begin{tabular}{c|ccccccc}
\hline\hline
      \begin{minipage}{1.9cm}  \end{minipage}&
      \begin{minipage}{0.5cm} $\chi$ \end{minipage}
      &\begin{minipage}{1cm}$20$ \end{minipage} &
      \begin{minipage}{1cm} $40$ \end{minipage} &
      \begin{minipage}{1cm} $60$ \end{minipage} &
      \begin{minipage}{1cm} $70$ \end{minipage} &
      \begin{minipage}{1cm} $78$ \end{minipage} \\
\hline
 \begin{minipage}{1.9cm}$\Delta=-0.9999$\end{minipage}
 & $d_f$& 0.799 & 0.897 & 0.9406 & 0.9504 & 0.953 \\
 \hline\hline
 & $\chi$ & $20$ & $40$ & $60$ & $80$ & $100$\\
\hline
 \begin{minipage}{1.9cm} $\Delta=-0.99999$\end{minipage}
 &  $d_f$& 0.7998 & 0.8992 & 0.9486 & 0.97 & 0.972\\
 \hline\hline
\end{tabular}
\label{table3-2}
\end{table}

\subsubsection*{2. The fractal dimension $d_f$ for the spin-1 bilinear-biquadratic model in the vicinity of the
$\rm{SU}(3)$ ferromagnetic point}
\begin{figure}
\includegraphics[width=0.48\textwidth]{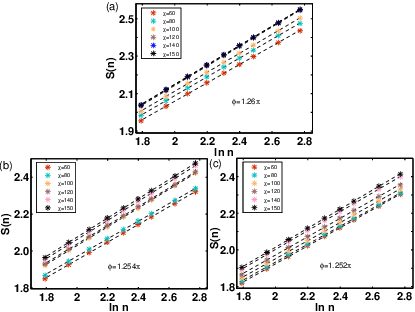}
\caption{(color online) The finite block-size scaling of the entanglement entropy $S(n)$ in the pseudo fractal regime for the spin-1
   bilinear-biquadratic model, with the bond dimension $\chi=60, 80, 100, 120, 140$ and $150$,
 for (a) $\phi=1.26\pi$; (b) $\phi=1.254\pi$; and (c) $\phi=1.252\pi$.
}
\label{BBfrac}
\end{figure}

The fractal dimension $d_f$ is 2 for the model (1) at the $\rm{SU}(3)$ ferromagnetic point~\cite{shiqq0}. We need to
extract the fractal dimension $d_f$ from the finite block-size scaling in the pseudo fractal regime in the vicinity of the
$\rm{SU}(3)$ ferromagnetic point for the spin-1 bilinear-biquadratic model.
In Fig.~\ref{BBfrac}, the finite block-size scaling of the entanglement entropy $S(n)$ is performed, with the bond dimension
$\chi=60, 80, 100, 120, 140$ and $150$, for (a) $\phi=1.26\pi$; (b) $\phi=1.254\pi$; and (c) $\phi=1.252\pi$, which are located
in the pseudo fractal regime.
The best linear fit yields that the fractal dimension $d_f=1$, with $n$ ranging from $6$ to $16$, with a relative error being less than $5\%$,
 as shown in Table~\ref{table4}. Note that the fractal dimension $d_f$ tends to saturation, as the bond dimension $\chi$ increases.

\begin{table}
\renewcommand\arraystretch{2}
\caption{The fractal dimension $d_f$ in the pseudo fractal regime is extracted from the $\rm{U}(1)$ iDMRG simulation
for the spin-1 bilinear-biquadratic model, with the block size $n$ being from $6$ to $16$.
Here, $\phi=1.26\pi, 1.254\pi$ and $1.252\pi$.}
\begin{tabular}{c|cccccccccc}
\hline\hline
       \begin{minipage}{1.5cm}  \end{minipage}&
      \begin{minipage}{0.5cm} $\chi$ \end{minipage}
      &\begin{minipage}{0.9cm}$60$ \end{minipage} &
      \begin{minipage}{0.9cm} $80$ \end{minipage} &
      \begin{minipage}{0.9cm} $100$ \end{minipage} &
      \begin{minipage}{0.9cm} $120$ \end{minipage} &
      \begin{minipage}{0.9cm} $140$ \end{minipage} &
      \begin{minipage}{0.9cm} $150$ \end{minipage} \\
\hline
 \begin{minipage}{1.5cm} $\phi=1.26\pi$ \end{minipage}
 &$d_f$& 0.9802 & 1.0024 & 1.0122 & 1.0384 & 1.035 & 1.0364 \\
 \begin{minipage}{1.5cm} $\phi=1.254\pi$ \end{minipage}
 &$d_f$& 0.9604 & 0.9554 & 1.0178 & 1.0148 & 1.0368 & 1.0418 \\
 \begin{minipage}{1.5cm} $\phi=1.252\pi$ \end{minipage}
 &$d_f$& 0.9874 & 0.9794 & 0.9898 & 1.0018 & 1.0314 & 1.0404 \\
 \hline\hline
\end{tabular}
\label{table4}
\end{table}

\begin{figure}
\includegraphics[width=0.35\textwidth]{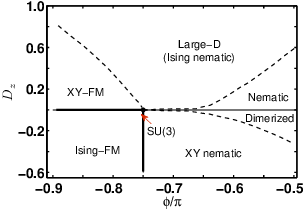}
\caption{(color online) A sketch of the ground state phase diagram close to the $\rm{SU}(3)$ ferromagnetic point for
the spin-1 bilinear-biquadratic model (\ref{BBD1}) undergoing a quadratic Zeeman effect. This is adapted from Ref.~\cite{BBDmodel0}}. \label{DFig}
\end{figure}

\section*{SVIII. The origins of the pseudo fractal and pseudo critical regimes in the tensor network simulations.}\label{Origins}

We have demonstrated in Sec. SIII that, in the pseudo critical regime,  pseudo SSB from $\rm{SU}(2)$ symmetry to $\rm{U}(1)$ occurs. Here,  $\rm{U}(1)$ is generated from $S^{\vec{n}}$, with the directional vector $\vec{n}$ being random and depending on
an initial state for a ground state wave function generated from the iTEBD simulation. Therefore, two generators are spontaneously broken, yielding two pseudo GMs of type-A. Therefore, the central charge $c$ takes  $c=2$, given that the central charge $c$ measures the number of gapless low-lying excitations~\cite{CFT0}. The same argument also applies to a ground state wave function generated from the $\rm{U}(1)$ iDMRG simulation, with the only difference being that spontaneous polarization is always along the $z$-axis in the spin space. This is in sharp contrast to the
$\rm{SU}(2)$ iDMRG simulation, in which the full $\rm{SU}(2)$ symmetry is implemented so that no pseudo SSB is allowed to occur. Thus, the pseudo critical regime only occurs in the iTEBD simulation and the $\rm{U}(1)$ iDMRG simulation, but not in the
$\rm{SU}(2)$ iDMRG simulation.

As for the pseudo fractal regime, it only occurs in the $\rm{U}(1)$ iDMRG simulation, since a ground state with the $z$-component of the total spin being zero is targeted, ensuring that it is highly entangled so that a scale invariant ground state, characterized in terms of the fractal dimension $d_f = 1$ corresponding to one pseudo GM of type-B, is able to compete with a ground state, characterized in terms of the central charge $c = 2$ corresponding to two pseudo GMs of type-A. In contrast, the $z$-component of the total spin is not preserved in the iTEBD simulation. Thus, it is impossible to produce such a highly entangled scale invariant ground state. We remark that the fractal dimension $d_f$ is equal to 1 in the pseudo fractal regime, since there is one pseudo GM of type-B as a result of pseudo SSB from $\rm{SU}(2)$ to $\rm{U}(1)$. This pseudo GM of type-B evolves into one of the two GMs at the $\rm{SU}(3)$ ferromagnetic point. Indeed,
at this point, two GMs of type-B emerge, as a consequence of SSB from $\rm{SU}(3)$ to $\rm{SU}(2) \otimes \rm{U}(1)$, implying that the fractal dimension $d_f$ is equal to 2. Here, we note that the fractal dimension $d_f$ measures the number of low-lying excitations~\cite{shiqq0} (also cf. Sec. SVII in SM).  As a consequence, one may conclude that the pseudo fractal regime originates from the proximity effect to a highly entangled scale invariant ground state at the $\rm{SU}(3)$ ferromagnetic point.

\begin{figure}
\includegraphics[width=0.33\textwidth]{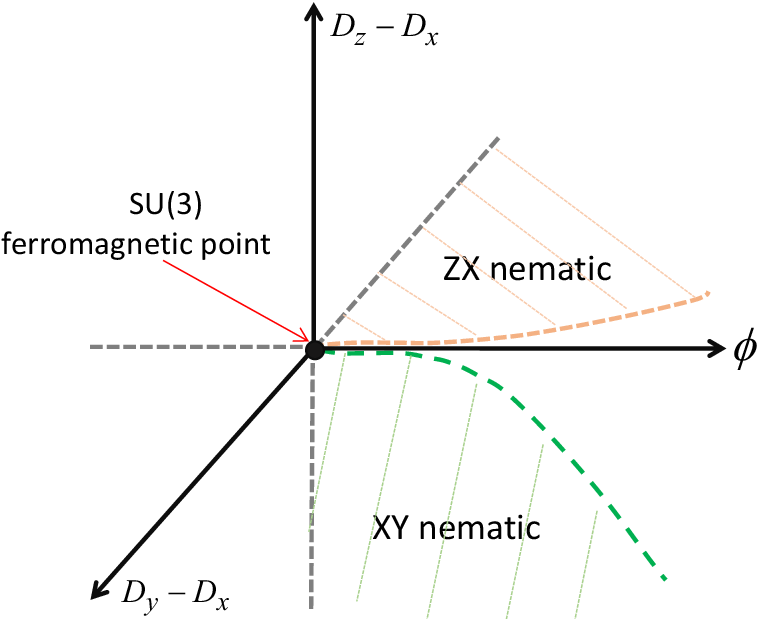}
\caption{(color online) A sketch of the two critical nematic phases, close to the $\rm{SU}(3)$ ferromagnetic point, for
the spin-1 bilinear-biquadratic model (\ref{BBD1}) undergoing a quadratic Zeeman effect, if one chooses $D_x$ to be a fixed constant. Note that the XY and ZX nematic critical phases are symmetric under the unitary transformation: $S_j^y \leftrightarrow S_j^z$ and $S_j^x \leftrightarrow - S_j^x$. We emphasize that the model possesses a $\rm{U}(1)$ symmetry only when $D_y=D_x$ or $D_z=D_x$, which is necessary for a critical nematic phase to emerge.} \label{DFig2}
\end{figure}

The final question we need to address concerns the origin of the pseudo critical regime. Given there is only one phase transition point from the ferromagnetic phase to the dimerized phase located at the $\rm{SU}(3)$ ferromagnetic point, the model (1) itself does not accommodate any critical  nematic phase close to the transition point. Therefore, one may expect that a critical nematic phase exists in the vicinity of the
$\rm{SU}(3)$ ferromagnetic point, if extra interactions are introduced into the model (1). Since
$\langle S_j^x \rangle = 0$, $\langle S_j^y \rangle = 0$, and $\langle S_j^z \rangle = 0$, in both the iTEBD simulation and the $\rm{U}(1)$ simulation, the linear Zeeman effect is not relevant. Instead, we turn to a quadratic Zeeman effect, described by the Hamiltonian:
\begin{equation}
\begin{aligned}
 H_{\rm{BBXYZ}} = &J\sum_{j}
 \left[(\cos\phi (\mathbf{S}_{j}\mathbf{S}_{j+1})
       +\sin\phi (\mathbf{S}_{j}\mathbf{S}_{j+1})^2)\right]+\\
       &D_x\sum_{j}(S_{j}^x)^2
       +D_y\sum_{j}(S_{j}^y)^2+D_z\sum_{j}(S_{j}^z)^2.
 \end{aligned}
 \label{BBD1}
\end{equation}
Here, $D_x$, $D_y$, and  $D_y$ are coupling constants describing a quadratic Zeeman effect. A special case with  $D_x=0$ and  $D_y=0$ has been investigated in Ref.~\cite{BBDmodel0}. The ground state phase diagram is sketched in Fig.~\ref{DFig}, close to the $\rm{SU}(3)$ ferromagnetic point. It is found that the XY nematic phase, with the central charge $c=1$, extends up to the
$\rm{SU}(3)$ ferromagnetic point. In addition, the phase transition from the XY nematic phase to the dimerized phase is of the Kosterlitz-Thouless type. Therefore,  for the model (\ref{BBD1}), there should be the XY and ZX critical nematic phases,
with the central charge $c=1$, in the vicinity of the $\rm{SU}(3)$ ferromagnetic point, due to the local constraints: $(S_{j}^x)^2   +(S_{j}^y)^2+(S_{j}^y)^2 =2$, if one chooses $D_x$ to be a fixed constant. The two critical nematic phases are sketched in Fig.~\ref{DFig2},
which are symmetric under the unitary transformation: $S_j^y \leftrightarrow S_j^z$ and $S_j^x \leftrightarrow - S_j^x$.
We emphasize that no other critical regime with the central charge $c=1$ exists in the vicinity of the $\rm{SU}(3)$ ferromagnetic point,
since the model (\ref{BBD1}) possesses a $\rm{U}(1)$ symmetry only when $D_y=D_x$ or $D_z=D_x$, which is necessary for a critical nematic phase to emerge.
This explains why two pseudo GMs of type-A emerge in the iTEBD  and $\rm{U}(1)$ iDMRG simulations, as long as it is close enough to the $\rm{SU}(3)$ ferromagnetic point. In other words, it is the proximity effect to the two critical nematic phases with $c=1$, which are infinitesimally close to each other in the vicinity of the $\rm{SU}(3)$ ferromagnetic point, that constitutes the origin of the pseudo critical regime with $c=2$ in the iTEBD  and $\rm{U}(1)$ iDMRG simulations. However, the two critical nematic phases with $c=1$ do not meet each other at the $\rm{SU}(3)$ ferromagnetic point, though asymptotically close, implying that no  $\rm{SU}(2)$ WZW model with level $k=4$ exists as a limit of the pseudo critical regime, as the bond dimension $\chi$ tends to infinity.

Finally, the proximity effect to the two critical nematic phases with $c=1$ offers a mechanism to account for the opening of an exponentially decaying small gap, when the $\rm{SU}(3)$ ferromagnetic point is approached from the dimerized phase. This is due to the facts that the phase transition from the XY/ZX nematic phase to the dimerized phase is of the Kosterlitz-Thouless type, meaning an essential singularity arising from a marginally relevant perturbation at the Kosterlitz-Thouless transition points, and that the two lines of the critical points themselves get close to each other in the vicinity of the $\rm{SU}(3)$ ferromagnetic point.

We thank Murray Batchelor, Sam Young Cho, John Ove Fj{\ae}restad, Javier Rodr\'{i}guez-Laguna, Silvia N. Santalla, and Germ\'{a}n Sierra for enlightening discussions.
The work is supported by the National Natural Science Foundation of China (Grant No. 11805285).